\documentclass[12pt,preprint]{aastex}

\lefthead{Skinner et al.}
\righthead{Lynds 1228}

\begin{document}

\title{An  X-ray and Infrared Survey of the Lynds 1228 Cloud Core}

\author{Stephen L. Skinner\footnote{CASA, Univ. of Colorado,
Boulder, CO, USA 80309-0389; stephen.skinner@colorado.edu},
Luisa Rebull\footnote{Spitzer Science Center / Caltech, M/S 220-6, 1200 East California Blvd.,
        Pasadena, CA 91125; ~~~~~~~~~~~~~~~ 
        rebull@ipac.caltech.edu}, and
Manuel  G\"{u}del\footnote{Dept. of Astrophysics, Univ. of Vienna,
T\"{u}rkenschanzstr. 17,  A-1180 Vienna, Austria; manuel.guedel@univie.ac.at}}
%
\newcommand{\ltsimeq}{\raisebox{-0.6ex}{$\,\stackrel{\raisebox{-.2ex}%
{$\textstyle<$}}{\sim}\,$}}
%
\newcommand{\gtsimeq}{\raisebox{-0.6ex}{$\,\stackrel{\raisebox{-.2ex}%
{$\textstyle>$}}{\sim}\,$}}
%
\newcommand{\ai}{$A_I$}
\newcommand{\av}{$A_V$}
\newcommand{\cf}{{\rm cf.}}
\newcommand{\eg}{{\it e.g.}}
\newcommand{\etal}{et~al.}
\newcommand{\hk}{$H-K$}
\newcommand{\hks}{$H-K_{\rm s}$}
\newcommand{\ic}{$I_{\rm C}$}
\newcommand{\icks}{$I_{\rm C}-K_{\rm s}$}
\newcommand{\ie}{{\it i.e.}}
\newcommand{\ik}{$I-K$}
\newcommand{\iks}{$I-K_{\rm s}$}
\newcommand{\jhk}{$JHK_{\rm s}$}
\newcommand{\jk}{$J-K$}
\newcommand{\kh}{$K-H$}
\newcommand{\ks}{$K_{\rm s}$}
\newcommand{\lbol}{$L_{\rm bol}$}
\newcommand{\lsun}{$L_{\sun}$}
\newcommand{\lx}{$L_{\rm x}$}
\newcommand{\mdot}{$\dot{M}$}
\newcommand{\msun}{M$_{\sun}$}
\newcommand{\rc}{$R_{\rm C}$}
\newcommand{\ri}{$R-I_{\rm C}$}
\newcommand{\rsun}{R$_{\sun}$}
\newcommand{\teff}{$T_{\rm eff}$}
\newcommand{\ubvri}{$UBVR_{\rm C}I_{\rm C}$}
\newcommand{\uv}{$U-V$}
\newcommand{\vi}{$V-I$}
\newcommand{\vsini}{$v \sin i$}
\newcommand{\mum}{$\mu$m}
%

\begin{abstract}
The nearby Lynds 1228 (L1228) dark cloud at a distance of ~200 pc is known
to harbor several young stars including the driving sources of the giant 
HH 199 and HH 200 Herbig-Haro outflows. L1228 has been previously studied
at optical, infrared, and radio wavelengths but not in X-rays. We present
results of a sensitive 37 ks Chandra ACIS-I X-ray observation of the 
L1228 core region. Chandra detected 60 X-ray sources, most of which 
are faint (<40 counts) and non-variable. Infrared counterparts were 
identified for 53 of the 60 X-ray sources using archival data from 
2MASS, Spitzer, and  WISE.  Object classes were assigned 
using mid-IR colors for those objects with complete photometry, most of which
were found to have colors consistent with extragalactic
background sources. Seven young stellar object (YSO) candidates were identified including
the class I protostar HH 200-IRS which was detected as a faint
hard X-ray source. No X-ray emission was detected from the luminous
protostar HH 199-IRS. 
We summarize the X-ray and infrared properties of
the detected sources and provide IR spectral energy distribution modeling
of high-interest objects including the protostars driving the HH outflows.
\end{abstract}



\section{Introduction}
Lynds 1228 (hereafter L1228; centered at J2000 RA = 20.97$h$, Dec. = $+$77.32$^{\circ}$) is one 
of the closest  dark clouds in the  catalog of Lynds (1962). It is part of the 
Cepheus Flare region  which in its entirety comprises  one of the nearest giant molecular clouds.
The Cepheus Flare  consists of as many as eight different dark cloud complexes widely distributed over 
$\sim$100 deg$^{2}$ (Kun 1998; Kiss et al. 2006). 
The main cloud complex lies at d = 300 $\pm$ 30 pc but L1228 is
in the foreground at d = 200$^{+100}_{-20}$ pc (Kun 1998; Kun, Kiss, \& Balog 2008).
The total mass of L1228 is estimated to be M$_{cloud}$ $\sim$ 230 M$_{\odot}$
which is intermediate within the range $\sim$40 - 1500 M$_{\odot}$ for the 
other cloud complexes in the Cepheus Flare (Kirk et al. 2009). 

Star-formation has occurred in several of the clouds as evidenced by
the discovery of a number of low-mass young stellar objects (YSOs)
in various surveys, as summarized below. 
But several dense cores are also present which appear starless and 
may eventually form low-mass stars (Kun 1998). Star-formation in some of
the clouds may have been triggered by a supernova which exploded 
$\sim$40,000 years ago whose vestige is an  expanding  hot bubble    
known as the Cepheus Flare Shell (Grenier et al. 1989; Olano, Meschin, \& Niemela 2006).
L1228 is of particular interest since its projected position lies just inside
the current radius of this expanding shell (Fig. 1 of Kirk et al. 2009).
As such, it emerges as a likely candidate for recent star-formation.

An objective prism survey of L1228  was undertaken by  Ogura \& Sato (1990) to identify
candidate pre-main sequence (PMS) stars. Their survey identified
69 H$\alpha$ emission stars of which nine clustered near the central part of L1228.
Using optical spectroscopy and photometry in combination with Two-Micron All-Sky 
Survey (2MASS) near-IR
and {\em Spitzer} Space Telescope  IRAC and MIPS-24 $\mu$m data, Kun et al. (2009) identified 
13 PMS stars in L1228 with extinctions in the range A$_{\rm V}$ = 0.0 - 5.9 mag
(median  A$_{\rm V}$ $\approx$ 3 mag) and ages spanning 1 - 10 Myr with a median
of $\sim$3 My.   The {\em Spitzer} Gould Belt Survey (Kirk et al. 2009) included
L1228 in which 23  YSO candidates were identified. 
Further analysis of JHK photometry combined
with {\em Spitzer} IRAC and MIPS 24$\mu$m/70$\mu$m data 
by Chapman \& Mundy (2009) identified seven YSOs in the L1228 core region.
The YSOs identified so far are predominantly class II objects (e.g. Fig. 8 of 
Kirk et al. 2009), also known as classical T Tauri stars (cTTS). These objects
are low-mass PMS stars still surrounded by accretion disks. However, at least
two young class I protostars are also present, HH 199-IRS and 
HH 200-IRS. High-mass stars such as the young OB stars found in
Orion are not currently present and the Cepheus Flare complex thus more
closely resembles regions such as Taurus where low-mass stars
are the rule.

The studies cited  above have focused on optical and IR identification of candidate
YSOs.  X-ray observations are also useful for this purpose since low-mass 
YSOs typically show elevated levels of X-ray emission relative to their main sequence
counterparts. L1228 has not previously been observed in a pointed X-ray observation
but it was part of the {\em ROSAT} All-Sky Survey (RASS) coverage region in 
Cepheus studied by Tachihara et al. (2005). They identified 16  
weak-lined T Tauri stars (wTTS) in  Cepheus but none lie in L1228.
We present here the results of a  more sensitive X-ray observation of the L1228
cloud core with {\em Chandra}. Our goals are to obtain the first census of
X-ray sources in the cloud core and  identify potential X-ray emitting YSOs.
Because of the proximity of  L1228  and
its relatively low extinction, the observation provides high sensitivity 
and is capable  of detecting X-ray emitting T Tauri stars  down to the
low end of the stellar mass limit (Sec. 2). 
We also analyze IR data from 2MASS, {\em Spitzer}, and the Wide-field Infrared
Survey Explorer  ({\em WISE}) in order to identify counterparts to X-ray sources, 
distinguish between YSOs and extragalactic contaminants, and determine
basic YSO properties. The study
goes beyond  previous work by including new X-ray data and by incorporating 
all previous {\em Spitzer} archive data as well as new {\em WISE} mid-IR data.

\section{Observations and Data Reduction}

\subsection{Chandra}

The {\em Chandra} observation is summarized in Table 1.
The exposure was obtained using the ACIS-I (Advanced CCD
Imaging Spectrometer) imaging array in faint timed-event mode.
ACIS-I has a combined field of view (FoV) of $\approx$16.9$'$ x 16.9$'$
consisting of four front-illuminated 1024 x 1024 pixel CCDs with a pixel size
of 0.492$''$. Approximately 90\% of the encircled energy at 1.49 keV
lies within $\approx$1$''$ of the central pixel of an on-axis  point source.
More information on {\em Chandra} 
and its instrumentation can be found in the {\em Chandra} Proposer's
Observatory Guide (POG)\footnote {See http://asc.harvard.edu/proposer/POG}.

The Level 1 events file provided by the {\em Chandra} X-ray
Center (CXC) was  processed using CIAO
version 4.4\footnote{Further information on
{\em Chandra} Interactive
Analysis of Observations (CIAO) software can be found at
http://asc.harvard.edu/ciao .}  standard science
threads. The  CIAO processing script {\em chandra\_repro} applied 
recent calibration updates. We used the CIAO {\em wavdetect} tool to
identify X-ray sources on the ACIS-I array along with their
centroid positions and 3$\sigma$ position error ellipses.
{\em Wavdetect} was executed on full-resolution images using events in the 
0.3 - 7 keV range to reduce the background. Wavelet radii as set by
the $scales$ input parameter were 
1, 2, 4, 8, and 16 pixels. 
Background was very low, amounting to 0.0104 counts per pixel (0.3 - 7 keV)
integrated over the full 36.8 ks exposure. A nominal on-axis source extraction
region of radius $r$ = 1$''$ contains only $\approx$0.14 background counts (0.3 - 7 keV).
We retained only those X-ray sources found by {\em wavdetect}
having $\geq$5 net counts in our final source list. This threshold
corresponds to an unabsorbed X-ray luminosity detection limit 
L$_{x}$ $\approx$ 10$^{28.05}$ ergs s$^{-1}$ (0.3 - 7 keV) at d = 200 pc, 
assuming a generic T Tauri star
thermal X-ray spectrum with plasma temperature kT = 3 keV and
hydrogen absorption column density  N$_{\rm H}$ = 10$^{21.8}$ cm$^{-2}$
(A$_{\rm V}$ $\approx$ 3 mag). This extinction is typical for 
the cloud as a whole (Kun et al. 2009) but A$_{\rm V}$ is higher
toward the most obscured central regions (Kiss et al. 2006; Chapman \& Mundy 2009;
Kirk et al. 2009). Using the known correlations between
L$_{x}$ and stellar mass (M$_{*}$) for TTS as determined from deep 
{\em Chandra} survey of the Orion Nebula Cluster (Preibisch et al. 2005),
and from the {\em XMM-Newton} survey of the Taurus Molecular 
Cloud (Telleschi et al. 2007),
the above L$_{x}$ limit corresponds to a TTS of mass 
M$_{*}$ $\ltsimeq$ 0.1 M$_{\odot}$.
The CIAO tool  {\em specextract} was used to extract X-ray
spectra of selected bright sources and spectral fitting was
carried out with HEASOFT XSPEC software vers. 12.4.0
\footnote{http://heasarc.gsfc.nasa.gov/docs/xanadu/xanadu.html}.                                  
Source variability probabilities were computed using the CIAO
tool {\em glvary} which is based on the Gregory-Loredo (GL)
algorithm (Gregory \& Loredo 1992, 1996).

\subsection{Infrared Archive Data (2MASS, Spitzer, WISE)}

We have used existing IR archive data to identify potential IR
counterparts to X-ray sources, obtain IR  photometry of high-interest
sources,  and determine object classes in those cases where good-quality
multi-band photometry are available.

\noindent {\em 2MASS:}~We obtained near-IR positions and magnitudes from  
the 2MASS All-Sky Catalog (Skrutskie \etal\ 2006).
2MASS provides  photometry at
J (1.25 $\mu$m), H (1.65 $\mu$m), and K$_{s}$ (2.17 $\mu$m).
The spatial resolution of $\approx$1.$''$2 is comparable to that of 
{\em Chandra} and to {\em Spitzer}  at 3.6 $\mu$m.

\noindent {\em Spitzer:}~
The Spitzer Space Telescope (Werner et al.  2004) observed the L1228
region multiple times with the Infrared Array Camera (IRAC; Fazio et al.  2004) and
the Multiband Imaging Photometer for Spitzer (MIPS; Rieke et al. 2004). IRAC 
provides data in four channels: I1 (3.6 $\mu$m), I2 (4.5 $\mu$m), I3 (5.8 $\mu$m),
and I4 (8.0 $\mu$m). The IRAC native pixel size is 1.$''$2 with spatial resolution
FWHM $\approx$ 1.$''$6 - 2.$''$0.  MIPS has three channels:
MIPS-24 (24 $\mu$m), MIPS-70 (70 $\mu$m), and MIPS-160 (160 $\mu$m)
with respective spatial resolutions FWHM $\approx$ 6$''$, 18$''$, and 40$''$.
Further information on these instruments can be found in the
respective {\em Instrument Handbooks}\footnote{http://irsa.ipac.caltech.edu/data/SPITZER/docs/}.
We have reprocessed all {\em Spitzer} IRAC and MIPS-24 archive data (Table 2)  intersecting
the {\em Chandra} FoV to generate deep-coverage image
mosaics. 
Since MIPS-70 and MIPS-160 band data are not
available for all the MIPS Astronomical Observation Requests (AORs)
we have only reprocessed the MIPS-24 data.
However, MIPS-70 photometry are available for some sources such as HH 199-IRS and HH 200-IRS  
from the Cores-to-Disks (c2d) {\em Spitzer} Legacy program (Evans et al. 2003).
We reprocessed the AORs starting from the  IRAC pipeline-corrected 
Basic Calibrated Data (BCD)  files or, for
MIPS photometry observations, the enhanced BCD (eBCD) files. 
For MIPS scan maps we started from the BCD files since
eBCD files are not available. We
used MOPEX (Makovoz \& Marleau 2005) to calculate overlap corrections
and to create exposure-time-weighted mosaics with a substantial  reduction in  
instrumental artifacts. Exposure time weighting is necessary since the total
integration time varies as a function of  position on the sky.
The pixel size for our mosaics was the same as the pipeline mosaics, 0.$''$6
for IRAC and 2.$''$45 for MIPS-24.
The {\em Spitzer} mosaics cover the full ACIS-I FoV
except for the NE and SW corners where 
2MASS and {\em WISE} provide  coverage.
We used
the IDL {\em aper.pro} routine to perform aperture photometry at the IR peak 
positions of sources of interest.
For the IRAC mosaics  we used an aperture radius $r$ = 3.$''$6 
and a background  annulus of $r$ = 3.$''$6 - 8.$''$4.
Aperture corrections in the four IRAC channels 
(1.124, 1.127, 1.143, \& 1.234 respectively) and IRAC
zero-point magnitudes (280.9, 179.7, 115.0, and 64.13 Jy, respectively)
were taken from the {\em IRAC Instrument Handbook}.
Point-response function (PRF) fitting of MIPS-24 sources was carried out
with MOPEX. In those cases where PRF fitting was not successful we obtained
MIPS-24 aperture photometry 
using an aperture of radius $r$ = 7$''$, an annulus of $r$ =  7$''$ - 13$''$, and an
aperture correction of 2.05 as per the {\em MIPS Instrument Handbook}.

\noindent {\em WISE:}~
We searched for mid-IR counterparts of X-ray sources using
the 14 March 2012 release of the {\em WISE} All-Sky Catalog (Wright et al. 2010).  
WISE completed an  all-sky survey  
during its cryogenic mission phase from January to August 2010.
WISE provides photometry
in four bands: W1 (3.4 $\mu$m), W2 (4.6 $\mu$m), W3 (12 $\mu$m),
and W4 (22 $\mu$m). Time variability can also be assessed based
on comparison of fluxes in each band for a given source obtained at different
times during the mission. WISE sensitivity is somewhat less than
obtained in our mosaiced {\em Spitzer} images of L1228
and WISE spatial resolution is modest with FWHM values of  6.1$''$,
6.4$''$, 6.5$''$, and 12.0$''$ in the W1, W2, W3, W4 bands.
WISE images and associated data in all four bands (e.g. instrumental
profile-fit photometry) were downloaded from the archive
for analysis.

\section{X-ray Results}

\subsection{Source Identification}

{\em Chandra} detected 60 X-ray sources on ACIS-I with
$\geq$5 net counts. Their positions on the  ACIS-I detector
are plotted in Figure 1 and source properties are summarized
in Table 3. Most of the X-ray sources are faint and only eight 
sources have $>$50 net counts. The mean number of counts is 25 and
the median is 13 counts. None of the X-ray sources
are listed  in the less sensitive RASS compilation of Tachihara et al. (2005).
We found potential  IR counterparts  for 53 of the  
60 sources (Table 4). In almost all  cases the offsets between
the X-ray and IR positions are $<$2$''$. As Table 4 shows, larger offsets 
of 2.$''$0 - 3.$''$6 are present for some sources (CXO sources  2, 5, 14, 16, 58, 60).
These  sources lie  off-axis and have large X-ray position error
ellipses due to broadening of the point-spread-function (PSF), so 
their  IR counterpart identifications  given in 
Table 4 are more uncertain.  WISE coverage spans the entire 
ACIS-I field-of-view and potential WISE counterparts were found
for 30 sources. Four WISE sources have variability flag values
$varflg \geq$ 6 in one or more bands indicating likely IR
variability (CXO 26, 27, 30, 53). Figure 2 is an overlay of the X-ray source
positions on a {\em WISE} W1-band image. The mosaiced {\em Spitzer}
images are more sensitive than {\em WISE} and {\em Spitzer}
counterparts were found for all but ten X-ray sources, seven of which 
lie outside the {\em Spitzer} coverage area (Table 4). IR photometry for 
each source are summarized in Table 5.

Interestingly, the brightest IR source in the field 
IRAS 20582$+$7724 = HH 199-IRS was not detected by {\em Chandra}.
It is a class I protostar  
that is believed to power the HH 199 outflow (Bally et al. 1995).
Its IR properties are discussed further in Section 5.
The hardest X-ray source as gauged by its mean photon energy
$\overline{\rm E}$ = 5.1 keV is CXO source 26 (J205706.72$+$773656.1).
This mean energy is well above the average value
$\overline{\rm E}$ = 2.8 keV  for the sample (Fig. 3).
The  X-ray position of CXO source 26 as determined by {\em wavdetect} is offset by 
only 0.$''$06 from the radio source L1228 VLA 4 (Reipurth et al. 2004) which is identified 
with the IR source HH 200-IRS that is  thought to power the HH 200 outflow.
As discussed further in Section 5, a second IR source is located 5.$''$7 northeast
of HH-200 IRS but is well outside the X-ray position error ellipse    and  is thus  
ruled out as the X-ray source. We thus associate this faint hard X-ray 
source with  HH 200-IRS, a known class I protostar (Chapman \& Mundy 2009).
Other previously known YSOs
detected by {\em Chandra} are CXO sources 30, 36, and 53 and their
properties are discussed further in Section 4.

\subsection{X-ray Variability}
X-ray variability is a common feature of YSOs
but only four L1228 sources (CXO 8, 10, 31, 50) show a high probability of
variability P$_{var}$ $\geq$ 0.9. They are all relatively
X-ray bright  with $>$50 net counts (Fig. 3).  This
suggests that variability could  be present in some fainter
sources but went undetected because of low count rates.
The X-ray light curves of the variable sources are shown in Figure 4.
In all cases the variability is rather low-level with 
count rates changing  by factors of $\sim$2 - 3.
No large-amplitude  X-ray flares were detected.

\subsection{X-ray Spectra}

The analysis of X-ray spectra can potentially provide
useful information on the line-of-sight neutral hydrogen
absorption column density (N$_{\rm H}$) toward the source
and characteristic plasma temperatures (kT) for thermal sources,
or photon power-law indices ($\Gamma$) for nonthermal sources such as active
galactic nuclei (AGN).
In our experience, at least $\sim$50 counts are needed to
justify rudimentary spectral analysis and we have thus
extracted spectra for the eight sources with $>$50 counts.
We attempted to fit the spectra with an absorbed  isothermal
(1T) thermal plasma model (the $apec$ model in XSPEC)
and also with an absorbed power-law (PL) model for those sources
which were classified as galaxies or AGNs based on their
IR properties (Sec. 4).
In all but two cases the simple 1T model gave
statistically acceptable  results with realistic
plasma temperatures of kT $\approx$0.8 - 6 keV.
The two exceptions (CXO 10 and 50) are discussed below.
The spectral fit results are summarized in Table 6.

The median of the  N$_{\rm H}$ values for the eight sources in Table 6 
is N$_{\rm H,med}$ = 5.25 $\times$ 10$^{21}$ cm$^{-2}$, where we  have
used the average of the two slightly different N$_{\rm H}$ values for those sources 
fitted with both a thermal and a power-law model. Using the conversion  
N$_{\rm H}$  = 2.2 $\times$ 10$^{21}$A$_{\rm V}$ cm$^{-2}$ 
of Gorenstein (1975), the median value is  A$_{\rm V,med}$ = 2.4 mag.
The conversion N$_{\rm H}$  = 1.6 $\times$ 10$^{21}$A$_{\rm V}$ cm$^{-2}$
of Vuong et al. (2003) gives  A$_{\rm V,med}$ = 3.3 mag.
Although the sample is small, the latter value 
A$_{\rm V,med}$ = 3.3 based on the Vuong et al. conversion 
is nearly identical to the median A$_{\rm V,med}$ = 3.1 obtained in 
the optical study of Kun et al. (2009).

The 1T model did not give an acceptable spectral fit
for CXO source 10 but a two-temperature (2T) model
was acceptable. It is the brightest
X-ray source in the sample and is variable. It is
classified as a star in the 
{\em HST} GSC (Table 3). Its X-ray variability is a sign of
magnetic activity and if it is a YSO then it is likely a
weak-lined T Tauri star (Sec. 4.5). Multi-temperature 
thermal plasma models are often required to fit the spectra
of X-ray bright T Tauri stars where higher signal-to-noise
ratio more tightly constrains the shape of the spectrum.
Thus, the need for a 2T model in the case of CXO 10
is not unusual and may simply reflect the  higher
quality of its spectrum.

CXO source 50 is also noteworthy. It is the second brightest
X-ray source and is classified as a AGN based on its 
{\em Spitzer} and {\em WISE} infrared properties (Sec. 4). 
But its X-ray emission is
likely variable (P$_{var}$ = 0.98) and X-ray variability  
is more commonly associated with magnetically-active YSOs
in star-forming regions. 
Both 1T and 2T thermal models require a high plasma temperature
component kT $\gtsimeq$ 20 keV.  Such a high temperature 
is well above the typical range kT $\sim$ 1 - 6 keV for 
YSOs but is not impossible for a flaring magnetically-active
YSO. If the IR classification as  a AGN is indeed correct, then the 
X-ray emission would more likely be nonthermal with  a power-law spectrum. 
We thus fitted the spectrum with
an absorbed power-law model and the fit is acceptable with 
a $\chi^2$ value nearly identical to that of the thermal model (Table 6).
We thus cannot distinguish between a hot thermal plasma spectrum
and a  power-law spectrum on the basis of fit statistics for CXO source 50.

\section{YSO Identification}

\subsection{Methods}

We have used multi-wavelength IR data to identify those X-ray
sources that are potential YSOs. Object classifications based on
IR analysis are given in  Table 3.
We have used IR color cuts (where color refers to the magnitude 
difference in two different bands) and color-magnitude constraints 
to identify candidate YSOs in the presence of contaminants 
such as star-forming galaxies with enhanced PAH emission, AGNs,
and shock emission knots. Several different color analysis methods
have been discussed in the literature and we have adopted 
that of  Gutermuth et al. (2008) for {\em Spitzer} data and 
Koenig et al. (2012) for WISE data. Similar IR analysis  for
the Taurus-Auriga  region based on these methods was
presented by Rebull et al. (2010, 2011).
We have restricted our
IR color analysis to those objects for which IR photometry 
is available in all four IRAC bands or all four WISE bands
(or both). In our sample, 41 objects
have 4-band photometry from IRAC or WISE (or  both),
and 10 of these also have 2MASS JHK$_{s}$ photometry (Table 5).

The color space methods implement cuts based on the photometric
properties of previously identified extragalactic sources, shock
emission features, and YSOs. It should be emphasized that classification
schemes based on IR color diagrams  are statistical in nature and are not 
100\% reliable.   Such schemes are tuned to finding IR-bright YSOs.
Contamination from AGNs becomes
significant for fainter IR sources and the contamination rate
increases as the objects become fainter. The schemes of Gutermuth et al. (2008)
and Koenig et al. (2012) thus also apply magnitude cuts  to
identify potential faint extragalactic sources. For example,
the Gutermuth  et al.  criteria require [I2] $>$ 13.5 mag 
along with other color constraints before an object is classified
as a broad-line AGN. Even so, low-mass
embedded YSOs can occupy these IR-faint regimes.
Closely-spaced sources can also
affect color analysis results, especially for lower resolution
MIPS and WISE data. We have inspected the IR images of all
{\em Chandra} sources and those with more than one IR source
within the {\em Chandra} position error ellipse are noted 
in Table 4.

IR color analysis is capable of distinguishing between heavily-reddened
class I protostars with infalling envelopes and  class II objects (i.e. cTTS 
with accretion disks).
Class III objects (weak-lined T Tauri stars) have little or no infrared
excess and are generally not identifiable on the basis of IR colors,
as demonstrated  below (e.g. CXO source 36).
Once a YSO candidate has been identified, its status as a 
class I, II, or III object  can also be discerned  by fitting the 
IR spectral energy distribution (SED) to determine the IR spectral index 
$\alpha$ = $d$~log($\lambda$F$_{\lambda}$)/$d$~log$\lambda$
where F$_{\lambda}$ is the flux density at wavelength $\lambda$.
In the approach of Haisch et al. (2001), $\alpha$ is 
evaluated in the $\approx$2 - 10 $\mu$m range.
Class I sources
have  $\alpha$ $>$ 0.3, flat spectrum sources have
$-$0.3 $\leq$ $\alpha$ $<$ 0.3, class II sources
have $-$1.6 $\leq$ $\alpha$ $<$ $-$0.3, and class III
sources have $\alpha$ $<$ $-$1.6. We have computed 
spectral indices for those sources suspected of being
YSOs for comparison with classes assigned on the basis
of IR colors.

\subsection{Extragalactic Contaminants}

Because of the modest extinction toward L1228 (median A$_{\rm V}$ 
$\approx$ 3 mag) and its location above the galactic plane
($b$ = $+$20.2$^{\circ}$), the X-ray sample is expected to contain
a substantial number of extragalactic sources. Indeed, of the 
41 X-ray sources with 4-band IRAC or {\em WISE} mid-IR photometry
suitable for color space analysis,  28 are classified as AGNs or 
star-forming galaxies. This yields an estimated extragalactic source 
fraction of 28/41 (68\%). There are undoubtedly other as yet 
unidentified extragalactic sources in the sample since twelve 
sources with IR counterparts lacked sufficient photometry for
color-based classification  and seven X-ray sources have no counterparts.
We also note here that nine X-ray sources having complete 4-band 
IR photometry did not pass any of the color-based criteria for 
classification as  AGN, PAH-emission star-forming galaxy, shock emission knot, 
or YSO. Optical catalog searches show that four of these are  stars
(CXO 10, 29, 36, 45; see Table 3) but the other
five objects remain unclassified (CXO 18, 28, 46, 49, 60).

All  eight of the brightest X-ray sources with  
$>$50 counts have 4-band mid-IR photometry and five of them are 
classified as extragalactic on the basis of their IR properties
(CXO 8, 31, 38, 50, 55).  We compare the above with the number
of bright  extragalactic X-ray sources ($>$50 counts) predicted
on the basis of source count data in the {\em Chandra} Deep Field-North 
(CDF-N) Survey as analyzed by Brandt et al. (2001). They show that the number
of X-ray sources $N$ per square degree with an {\em absorbed} hard-band (2 - 8 keV) 
flux greater than a specified lower limit F$_{x,hard}$  is 
$N$($>$F$_{x,hard}$) =
2820(F$_{x,hard}$/10$^{-15}$ ergs cm$^{-2}$ s$^{-1}$)$^{-1.0\pm0.3}$ deg$^{-2}$.
To apply this result, we need to know what value of F$_{x,hard}$ is
required to accumulate at least 50 counts (0.3 - 7 keV) from a generic
extragalactic source toward L1228 in a 36,779 s exposure. Note here that
we require $>$50 counts in the {\em broad} 0.3 - 7 keV band in order to make
comparisons with the L1228 net counts measurements in Table 3. Following
Brandt et al. we assume a power-law X-ray spectrum with a photon power-law index $\Gamma$ = 1.4
for the generic extragalactic source.  We  also
assume a median absorption column density toward L1228
of N$_{\rm H}$ = 5 $\times$ 10$^{21}$ cm$^{-2}$ (Sec. 3.1).
Under these spectral assumptions, the 
PIMMS\footnote{For more information on the Portable Interactive Multi-Mission Simulator (PIMMS)
see http://cxc.harvard.edu/ciao/ahelp/pimms.html . } 
simulator predicts an absorbed flux
F$_{x,hard}$ = 1.54 $\times$ 10$^{-14}$ ergs cm$^{-2}$ s$^{-1}$.
This value of F$_{x,hard}$ will yield 50 broad-band counts (0.3 - 7 keV)
and 27 hard-band counts (2 - 8 keV) in 36,779 s.
Substituting this value of F$_{x,hard}$  into the above expression for source
counts and scaling  the result to  the ACIS-I FoV
(0.07834 deg$^{2}$) gives  $N$($>$F$_{x,hard}$) = 14 (6 - 33)
where the range in parentheses reflects the uncertainties
in the CDF-N source count power-law fit index. Thus, the hard-band
CDF-N data predict at least six extragalactic sources in the
L1228 FoV with $>$50 broad-band counts and we have identified five candidates.
This small difference is not  problematic since we have assumed a
constant N$_{\rm H}$ toward L1228  but the absorption
is in fact higher toward the center of the cloud
(Kun et al. 2009; Chapman \& Mundy 2009). In addition, some
field-to-field variation in X-ray source counts is anticipated
(Cowie et al.  2002).

\subsection{YSO Candidates}

We have identified seven YSO candidates in the X-ray sample
(CXO sources 14, 16, 26, 30, 36, 53, 56). All of these
except CXO 36 were classified as YSOs on the basis of color
analysis using 4-band IRAC or {\em WISE} mid-IR photometry.
Representative color-color diagrams for these objects are shown 
in Figure 5. CXO source 36 was unclassified on the basis of
its IR colors but is designated a likely YSO by 
virtue of its  association  with H$\alpha$ emission star number 38 
(OSHA 38) in the list of Ogura \& Sato (1990). Our SED fits (Sec. 4.3) 
indicate that it has little if any IR excess so it is  probably
a class III object (wTTS).

Four of the YSOs are known from previous studies (CXO sources 26, 30, 36, and 53).
CXO 26 is associated with
the class I protostar HH 200 IRS  (Chapman and Mundy 2009).
CXO 30 was identified as a class II YSO (cTTS)
by Chapman and Mundy (2009) and our analysis confirms this.
CXO 36 is an H$\alpha$ emission-line star as  noted above.
CXO 53 is associated with the  emission-line star
OSHA 42 (Kun et al. 2009), a class II object which has
the strongest H$\alpha$ emission of the stars identified
in the survey of Ogura \& Sato (1990).

In addition to the above four previously-known YSOs, 
CXO sources 14, 16, and 56 are classified as
possible  YSOs on the basis of IR color analysis.
The YSO classifications of CXO 14 and 16 are questionable
as discussed further below,
but the classification of CXO 56 as a class II YSO
on the basis of both {\em Spitzer} and {\em WISE}
colors is relatively secure. The IR counterpart of
CXO 56 is offset by only 0.$''$48  from the X-ray
position, making its association with the X-ray
source highly probable. 

Both CXO 14 and 16 are faint in X-rays  and
the latter object lies near the north edge of the ACIS-I detector
(Fig. 1) where sensitivity is degraded and the PSF is broadened.
The {\em Spitzer} and {\em WISE} counterparts in Table 4  
are offset from the X-ray position of CXO 14  by  3.1$''$ and 3.6$''$ 
respectively. Similarly,  the {\em Spitzer} counterpart
of CXO 16 is offset by 3.6$''$ from its X-ray position.
Even though these offsets are
larger than for most of the other sources, the IR counterparts
do lie within the {\em Chandra} 3$\sigma$ position error ellipses.
The {\em Spitzer}   source located 3.1$''$  SE of CXO 14 is
resolved. This {\em Spitzer}   source has IR colors consistent
with a class I YSO but its corresponding {\em WISE} colors 
give a PAH-rich star-forming galaxy classification. 
Because of the above issues with mid-IR position offsets
and IR color-class  ambiguities,  the classification
of CXO sources 14 and 16 as YSOs will require further
confirmation. Tighter constraints on their X-ray positions could
be achieved by placing them more nearly on-axis where the 
{\em Chandra} PSF is sharpest and would provide a more 
reliable assessment of whether the {\em Spitzer} and {\em WISE} 
sources in Table 4 are legitimate counterparts.

\subsection{YSO Properties}

Four of the six objects identified as YSOs on the basis of
IR colors  have 2MASS near-IR photometry in addition
to {\em Spitzer} or WISE photometry in at least four bands.
Also, more recent near-IR photometry is available for
CXO sources 26 and 30 from Chapman \& Mundy (2009). 
We have modeled the IR SEDs of these  sources using
the online modeling tool of Robitaille et al. (2007).
The modeling results are summarized in Table 7. 
The fitting tool provides detailed information for
several different models listed in order of increasing
$\chi^2$ fit statistics. 
Substantial differences can be present for the value
of any fitted parameter between  different models.
In Table 7 we give the median value of each parameter
based on the  five models with the lowest 
$\chi^2$ values. The most noteworthy result from
the fits is the very high extinction toward
the class I protostar HH 200-IRS (CXO 26). Such  high absorption
would absorb most soft X-ray photons and may thus be largely
responsible for the low number of counts and high mean 
photon energy  detected by {\em Chandra}.

We have also fitted the IR SED of CXO 36 since it is a H$\alpha$ emission
star, even though it was not classified as a YSO on the basis of its IR colors.
The best-fit model gives no indication of a significant disk
and the mass of any remnant disk must be very low
(M$_{disk}$ $\ltsimeq$ 10$^{-8}$ M$_{\odot}$).
A pure stellar photosphere Kurucz model with T $\sim$ 3750 K 
provides a good fit. The presence of H$\alpha$ emission
without a significant disk is indicative of a class III
YSO (wTTS). This classification is further supported
by its SED power-law index $\alpha$ = $-$2.5 $\pm$ 0.1.

\subsection{Other Interesting Sources}

{\em Bright Variable X-ray Sources (CXO 8, 10, 31, 50)}:~
As already noted in Section 3.2, four of the brightest X-ray sources have a high
probability of variability P$_{var}$ $\geq$ 0.9. Three of these are classified as 
AGNs or galaxies on the basis of their IR colors (CXO sources 8, 31, 50).
No  counterparts were found for these three  sources
in a search of several galactic and extragalactic databases
including  the {\em HST} GSC v. 2.3.2, {\em Hipparcos}, and USNO B1 catalogs.
Nevertheless, since X-ray variability is a common trait of YSOs in star-forming regions,
deep follow-up observations and time-monitoring of these three sources 
would be worthwhile in order to confirm that they are indeed extragalactic
in nature.  The fourth variable X-ray source (CXO 10) is the brightest
X-ray detection and was unclassified on the basis of IR colors but
has an optical counterpart in the {\em HST} GSC (Table 3).
Its IR spectral index
is $\alpha$ = $-$2.45 $\pm$ 0.1 which would only be consistent
with a   class III object if it is a YSO.

{\em CXO 27}:~
This faint hard X-ray source was classified as a AGN on the
basis of both IRAC and {\em WISE} colors. It lies just
19$''$ south of HH 200-IRS (Fig. 7). Its IR emission may
be variable in the WISE W1 band ($varflg$ = 91nn).
Because of its possible IR variability and  projected 
position  close to other known YSOs, follow-up observations
of CXO 27 would be useful to determine if it is a YSO
masquerading as a AGN. If it is a YSO then its IRAC
spectral index $\alpha$ = $+$1.0 [$+$0.1 - $+$1.9; 90\% conf. range] 
would be consistent with either a class I or flat-spectrum source.

{\em CXO 29}:~
The bright X-ray source CXO 29 was  unclassified by
IR colors and it also has an optical counterpart in the {\em HST} GSC
classified as a star (Table 3). Its IR spectral index
$\alpha$ = $-$2.24 $\pm$ 0.13 would imply a class III object if
it is a YSO.

\section{Herbig-Haro Energy Sources}

\subsection{HH 199-IRS}

HH 199-IRS (= IRAS 20582$+$7724 = 2MASS J205712.94$+$773543.7) 
is the brightest infrared source in the 
L1228 cloud core and is believed to power the giant HH 199 
molecular outflow (Bally et al. 1995; Devine et al. 2009). 
This object was not detected by {\em Chandra} nor were any of
the optical emission knots in its outflow identified by Devine et al. (2009).
Nevertheless, because of its IR brightness and key role as  a HH driving 
source we summarize its IR properties. The discussion below is
restricted to 2MASS and WISE data since IRAC fluxes are 
affected by saturation.

Based on WISE colors (Fig. 5) it is classified as a
class I protostar by the Koenig et al. (2012) criteria.
A fit of its SED using 2MASS and WISE photometry gives
a spectral index $\alpha$ = $+$0.32 $\pm$ 0.32 (90\% confidence
errors). This again gives  class I  using
the criteria of Haisch et al. (2001) but the uncertainties
are large enough to allow a flat-spectrum source classification (i.e. borderline
case between class I and class II). No significant
IR variability was found by WISE ($varflg$ = 2214)
but it is worth noting that its outflow exhibits large
changes in morphology over time and may be precessing (Bally et al. 1995;
Devine et al. 2009).

We have modeled the  IR SED of HH 199-IRS using 
2MASS and WISE photometry (Table 5; Fig. 8).
If the interstellar extinction is loosely constrained
to be  A$_{\rm V,ism}$ $<$ 20 mag, then the
Robitaille et al. tool finds a good fit with a 
total extinction 
A$_{\rm V}$ = A$_{\rm V,ism}$ $+$ A$_{\rm V,cs}$ =
14.7 [14.5 - 16.1] mag, almost all of which is 
attributed to the interstellar component. 
The total system luminosity  is 
L$_{bol}$ = 7.0 [2.7 - 9.6] L$_{\odot}$ 
assuming a distance  d = 200 $\pm$ 20  pc.
These are the median values of the five best-fitting models
followed in brackets by the range of values. 
The above luminosity is at least an order of magnitude greater
than that of HH 200-IRS if both objects are at similar distances.
The inferred disk mass is M$_{disk}$ $\approx$
0.026 [0.022  - 0.032] M$_{\odot}$ with a disk accretion rate
$\dot{\rm M}_{disk}$ = 2.6 [1.5 - 7.2] $\times$
10$^{-7}$ M$_{\odot}$ yr$^{-1}$.
If we enforce a tighter constraint A$_{\rm V,ism}$ $<$ 13 mag
based on the maximum L1228 core extinction determined by
Kirk et al. (2009), then the modeling tool compensates
by adding in additional circumstellar absorption
but the resulting fit has  larger residuals than
the one summarized above. Thus, in terms of goodness-of-fit,
the models which invoke relatively high interstellar extinction
with low circumstellar extinction are favored.

\subsection{HH 200-IRS}

This source shows a nearly-flat IR SED in the 3 - 8 $\mu$m
range but it then rises steeply at longer wavelengths (Fig. 8).
WISE colors are consistent with a class I source (Fig. 5). 
Although its IR emission is  bright  it is only
faintly detected in X-rays. The X-ray emission is 
hard with all detected photons having energies
E $\geq$ 3.7 keV. The extinction toward this class I source
is  high. The five best-fit  SED models using the Robitaille
et al. tool (Table 7) yield a median interstellar extinction
A$_{\rm V,ism}$ = 5 [3 - 13] mag but an additional large  
circumstellar extinction is also required. We examined
several different models, all of which required 
A$_{\rm V,cs}$ $>$ 32 mag. This implies  that the central
protostar is heavily obscured by a surrounding envelope.
Any soft photons (E $<$ 2 keV)
emitted by the central source will thus be heavily absorbed
and undetected. Because of the high absorption, the intrinsic
X-ray properties of the central source are quite uncertain
but the presence of sufficient hard emission to penetrate
the dense circumstellar envelope is a strong clue that
magnetic processes are involved in the X-ray production.
Shocks created by impact of the outflow with 
surrounding material are expected to produce only soft
emission  (E $\ltsimeq$ 1 keV) which would be undetected unless 
viewed through lower absorption at large offsets from
the star. We have checked for faint X-ray emission at the
position of the optical emission knots in the HH 200-IRS
outflow listed in Table 1 of Devine et al. (2009)
and none was detected by {\em Chandra}.

A second IR source HH 200-IRSB (= 2MASS J205707.89$+$773659.7)
lies 5.$''$7 NE of HH 200-IRS. As Figure 6
shows, HH 200-IRSB is well outside the {\em Chandra}
position error ellipse and is not detected in X-rays.
Figure 7 is a {\em Spitzer} 3-color image of the region.
This second source
lies nearly on the HH 200  outflow axis, thus raising the
question of whether it might actually be driving the giant HH
outflow. This possibility cannot be totally ruled out but we believe
that HH 200-IRS is more likely to be the driving source because
of its elevated activity as evidenced by hard X-ray and radio
emission (Reipurth et al. 2004).

Figure 8 compares the SEDs of HH 200-IRS and the close
IR companion. The pair is resolved by 2MASS and IRAC but not
by MIPS or WISE. Overall, the SEDs of the two sources look
quite similar and both are likely in the protostellar stage.
As Figure 8 shows, there is a discrepancy 
between the JHK$_{s}$ fluxes of HH 200-IRS between 2MASS
and the more recent values obtained by Chapman \& Mundy (2009) using
the KPNO 4m. The 4m fluxes are a factor of $\sim$4 lower than
2MASS at J band and a factor of $\sim$2 lower at H and K$_{s}$.
This difference may be largely due to the better spatial
resolution of the KPNO 4m but some IR variability may also
be present. The {\em WISE} variability flag has values
$varflg$ = 6631 where a  value of 6 in both W1 and W2 bands
means that the source is potentially variable with
small amplitudes. 

\subsection{HH 199-IRS and HH 200-IRS in Comparison}

Our SED models indicate that L$_{bol}$ for HH 199-IRS
is about ten times greater than HH 200-IRS, in agreement with 
other studies (Chapman \& Mundy 2009; Kirk et al. 2009).
The extinction toward the two protostars determined from 
SED fits is high and  models require substantial circumstellar 
absorption in the case of HH 200-IRS.
Faint X-ray emission was detected from HH 200-IRS
but HH 199-IRS was undetected. Differences in 
the column density of X-ray absorbing gas 
toward the two protostars or their intrinsic X-ray
properties could account for this.
The high median photon energy 
of the events detected from HH 200-IRS is  a clear sign that its 
intrinsic X-ray spectrum contains a hard component,
undoubtedly of magnetic origin. X-ray flares in class I 
protostars can produce such hard emission and in some
cases the protostar only becomes detectable during an
X-ray flare (Imanishi, Koyama, \& Tsuboi 2001).
If HH 200-IRS was in an active flare state during our
{\em Chandra} observation then that could explain 
why it was detected as a hard source but HH 199-IRS
was not. Variability is expected during an X-ray flare
but we  lack sufficient counts in HH 200-IRS  to obtain 
a definitive test for X-ray variability, but IR variability 
is suspected as mentioned above.

\section{The Nature of the L1228 Core X-ray Population}

The {\em Chandra} observation reported here
comprises the first deep X-ray census of the
L1228 cloud core. The area observed by 
{\em Chandra} covers a relatively small 
0.28$^{\circ}$ $\times$ 0.28$^{\circ}$ field-of-view  
and thus does not capture the entire L1228 cloud
for which the region of highest extinction spans
roughly 1 deg$^{2}$ (Fig. 9 of Kun et al. 2009).

A  large fraction of the 
60 X-ray sources appear to be extragalactic
which is not surprising given that L1228 lies
well above the Galactic plane ($b$ = $+$20.2$^{\circ}$) 
and the cloud opacity is low (Lynds 1962). We have  identified 
7 known or suspected YSOs. This 
12\% YSO fraction should be interpreted as a 
lower limit. We lack sufficient information for 19 X-ray
sources to assess their YSO status, either
because no optical or IR counterpart was found
or because IR/optical  data were insufficient to 
assign an object class. 

It is of some interest that very few class III
sources (wTTS) have so far been identified in L1228,
but some have been found in other dark clouds in the
Cepheus Flare and in off-cloud regions 
(Tachihara et al. 2005; Kirk et al. 2009).
Since wTTS have weak H$\alpha$ emission by definition and lack
significant IR excesses they can be overlooked in
H$\alpha$ and infrared surveys. Of the 23 YSO candidates
in L1228 identified in the IR study of Kirk et al. (2009),
only one is classified as a wTTS. That object
(Kirk et al. nr. 136) lies outside the {\em Chandra}
FoV but was not reported as a RASS X-ray
detection by Tachihara et al. (2005). Is the wTTS population 
in L1228 truly absent or simply not yet seen? Some of the  X-ray sources
in our sample which lack IR color classifications may be
wTTS, two of which have already been mentioned 
(CXO 10 and 29). Follow-up observations to search
for stellar counterparts and spectral signatures 
of youth such as Li will be needed to determine
whether any of the unclassified X-ray sources
are indeed wTTS.

Another point worth noting is  that the X-ray luminosities of the 
L1228 YSO candidates are quite low. The seven YSO 
candidates have a median of 12 net counts with a
range spanning  5 - 75 net counts. Assuming a generic TTS
thermal X-ray spectrum with kT $\sim$ 3 keV and 
A$_{\rm V}$ $\approx$ 3 mag (Kun et al. 2009), this corresponds
to unabsorbed luminosity log L$_{x}$(0.3 - 7 keV) = 
28.43 [28.05 - 29.23] ergs s$^{-1}$ at the nominal
L1228 distance of 200 pc. These values increase by
0.35 dex at the high end of the  distance
range (d = 300 pc; Kun 1998) and by about 0.17 dex for the
higher mean extinction A$_{\rm V}$ $\approx$ 7 mag
obtained by Kirk et al. (2009). 
Based on the known correlations between L$_{x}$
and M$_{*}$ for TTS in Orion and Taurus, 
the above L$_{x}$ range corresponds to lower mass
TTS with  M$_{*}$ $\ltsimeq$ 0.7  M$_{\odot}$.
(Fig. 3 of Preibisch et al. 2005; Fig. 1 of 
Telleschi et al. 2007). Thus, if a similar 
L$_{x}$ $\propto$ M$_{*}$ correlation holds
for L1228,  we are  detecting a highly-dispersed
low-mass population of PMS stars that apparently 
is deficient in the more X-ray luminous TTS
with masses M$_{*}$ $\gtsimeq$ 1  M$_{\odot}$.
This conclusion would also apply to any of the
unclassified X-ray sources that may turn out to
be YSOs. They have  6 - 37 net counts
or log L$_{x}$ = 28.13 - 28.92 ergs s$^{-1}$ 
(at d = 200 pc, A$_{\rm V}$ = 3 mag, kT = 3 keV).

Some additional support for the dearth of
higher mass TTS  comes from the optical study
of L1228 by Kun et al. (2009). Of the 13 PMS stars they
identified in L1228, only one was more massive
than 1 M$_{\odot}$ and the median mass was
0.5 M$_{\odot}$.  One of the stars in their
sample, the class II source OSHA 42, was the
brightest X-ray source classified as a YSO in our  
sample. Kun et al. classify 
OSHA 42 as spectral type M0 with an estimated mass 
M$_{*}$ = 0.6 M$_{\odot}$ and L$_{bol}$ = 0.49 L$_{\odot}$. 
The value we obtain for L$_{bol}$ from IR SED modeling
is nearly identical (Table 7). To the extent that L$_{x}$ 
correlates with M$_{*}$ in L1228, it thus seems likely
that the remaining L1228 X-ray sources identified as
TTS (class II or possible III) with fainter emission than
OSHA 42 are lower mass objects.

\section{Summary}

The main results of our combined X-ray/infrared study of L1228
are summarized below.

\begin{enumerate}

\item {\em Chandra} has provided the first sensitive pointed
      X-ray observation of the Lynds 1228 cloud core. Previous 
      optical, IR, and radio  studies have shown that L1228 contains
      a modest population of low-mass PMS stars and the heavily-reddened
      protostars HH 199-IRS and HH 200-IRS. {\em Chandra} detected 
      60 X-ray sources, most of which are faint ($<$40 counts)
     and non-variable. 

\item Using reprocessed {\em Spitzer} archive data, along with 2MASS near-IR
      data and recent WISE All-Sky Survey mid-IR data, IR counterparts were
      found for 53 sources. Source classification based on IR colors
      and magnitudes was undertaken on
      41 of these for which photometry in at least four
      IRAC or WISE bands was available. 

\item Seven {\em Chandra} sources are classified as  YSOs,
      four of which were already known from previous studies.
      In addition, three  X-ray sources have optical counterparts
      in the {\em HST} GSC or Tycho  catalogs that are classified
      as stars but it remains to be determined whether they are PMS stars.
      We have classified  28 of the X-ray sources 
      as extragalactic on the basis of their IR properties
      but additional  extragalactic contaminants are 
      likely present among the  19 X-ray sources which lack 
      sufficient IR data to assign object classes. But some of
      the unclassified X-ray sources may be weak-lined T Tauri stars,
      which are conspicuously absent in previous YSO surveys
      of L1228. Evidence that some wTTS may be present
      comes from the X-ray detection of H$\alpha$ emission-line
      star OSHA 38 whose IR SED shows no evidence for a significant
      disk.

\item Disk properties were inferred for four of the YSOs having
      good-quality  IR photometry using the  SED
      modeling tool of Robitaille et al. (2007). Similar SED
      modeling was undertaken for the protostars HH 199-IRS and HH 200-IRS.

\item Faint hard X-ray emission was detected from HH 200 IRS,
      the suspected driving source of the HH 200 giant outflow.
      Its IR colors confirm that it is a class I protostar and
      there are hints that its IR emission is variable.
      A close IR companion located 5.$''$7 away  also has a rising
      IR SED characteristic of a protostar but was undetected by {\em Chandra}.
      A very high circumstellar extinction is inferred for HH 200-IRS 
      from  SED modeling, which likely accounts for its  
      hard X-ray emission. No X-ray emission was detected from
      HH 199-IRS but its IR colors and spectral index yield a
      a class I protostar classification. Modeling of its IR
      SED shows that it is at least an order of magnitude more
      luminous than HH 200-IRS if it lies at the same distance. 

\item The X-ray luminosities of the T Tauri stars  in L1228 detected by
      {\em Chandra} are low:  log L$_{x}$ $\ltsimeq$ 29.7 ergs s$^{-1}$,
      assuming A$_{\rm V}$ and distances at the high end of the 
      L1228 range. If a  correlation between L$_{x}$ and stellar
      mass exists among TTS in L1228 similar to that  in
      Taurus and Orion, then the inferred masses of the X-ray
      detected TTS are M$_{*}$ $\ltsimeq$ 0.7 M$_{\odot}$.
      Although this result is based only on partial X-ray
      coverage of the L1228 dark cloud, it is nevertheless consistent 
      with  the broader picture that L1228  has formed only low-mass stars.

\end{enumerate}

\acknowledgments

The scientific results reported in this article are based in part
on observations made by the {\em Chandra} X-ray Observatory.
We thank Kimberly Sokal and Bryan Nagel for assistance with 
data reduction.
This work is based in part on archival data obtained with the {\em Spitzer} 
Space Telescope, which is operated by the Jet
Propulsion Laboratory (JPL), California Institute of 
Technology (Caltech) under a  contract with NASA.
This publication makes use of data products from the Wide-field
Infrared Survey Explorer (WISE), which is a joint project of the University
of California, Los Angeles, and JPL/Caltech,  funded by NASA.
This research
has made use of data products from the Two Micron All-Sky Survey
(2MASS), which is a joint project of the University of Massachusetts
and the Infrared Processing and Analysis Center (IPAC), funded by 
NASA and the National Science Foundation.  These data were served by 
the NASA/IPAC Infrared Science
Archive, which is operated by JPL/Caltech under contract with 
NASA.

\clearpage

\begin{deluxetable}{ll}
\tabletypesize{\small}
\tablewidth{0pc}
\tablecaption{Chandra Observation of L1228  }
\tablehead{
\colhead{Parameter} &
\colhead{} \\
}
\startdata
ObsId                & 7425                     \nl
Start Date/Time (TT) & 2007-07-22/20:33:39      \nl
Stop  Date/Time (TT) & 2007-07-23/07:13:40      \nl
Instrument           & ACIS-I    \nl                                                           
Field-of-View (arcmin)\tablenotemark{a}        & 16.9 $\times$ 16.9  \nl
Livetime (s)         & 36,779                   \nl
Frame time (s)       & 3.2                      \nl
\enddata
\tablenotetext{a}{Nominal pointing position was at
(J2000) R.A. = 20$h$ 57$m$ 02.71$s$, 
Dec. = $+$77$^{\circ}$ 35$'$ 45.1$''$,
or ($l$,$b$) = (111.664$^{\circ}$, $+$20.226$^{\circ}$). }
\end{deluxetable}
\clearpage


\begin{deluxetable}{lllllll}
\tablecaption{Spitzer Observations of L1228}
\tablewidth{0pt}
\tabletypesize{\footnotesize}
\tablehead{
\colhead{Instrument}  &
\colhead{Mode}  &
\colhead{Program (PI)}  &
\colhead{AORKEY\tablenotemark{a}}  &
\colhead{Frame time} &
\colhead{Observation} &
\colhead{Notes}\\
 & & & & \colhead{(s)} &
 \colhead{date} & }
\startdata
IRAC & Map & 104 (Soifer) & 6577664 & 12\tablenotemark{b} &
2003-12-06 & First Look Survey (FLS)\\
IRAC & Map & 139 (Evans)  & 5161216 & 12\tablenotemark{b} & 2003-12-02 &
Cores-to-Disks legacy program \\
IRAC & Map & 3656 (Mundy) & 11400192 & 30  & 2004-12-15& \\
IRAC & Map & 3656 (Mundy) & 11391232 & 30  & 2004-11-27 &\\
MIPS & Scan & 104 (Soifer) & 6577408 & 2.6 & 2003-12-08 & FLS; 
MIPS 24, 70, 160 $\mu$m \\
MIPS & Photom. & 139 (Evans) & 5161472 & 2.6 & 2003-12-13 &very small footprint \\
MIPS & Photom. & 3656 (Mundy) & 11397632 & 9.96 & 2004-12-26 & MIPS 24 $\mu$m only \\
MIPS & Photom. & 3656 (Mundy) & 11395072 & 9.96 & 2004-12-07 & MIPS 24 $\mu$m only \\
\enddata
\tablenotetext{a}{An AOR (Astronomical Observation Request) is the fundamental unit of
Spitzer observing. An AORKEY is the unique 8-digit indentifier for the AOR which can be
used to retrieve these data from the Spitzer archive.}
\tablenotetext{b}{Obtained in high dynamic range (HDR) mode, consisting of a 
short 0.6 s and long 12 s frame  at each pointing.}
\end{deluxetable}

\clearpage

\begin{deluxetable}{lllllllll}
\tabletypesize{\scriptsize}
\tablewidth{0pt}
\tablecaption{Chandra X-ray Sources in  Lynds 1228\tablenotemark{a}}
\tablehead{
\colhead{CXO}      &
\colhead{R.A.}     &
\colhead{Decl.}    & 
\colhead{Net Counts}   &
\colhead{$\overline{\rm E}$}  &
\colhead{P$_{var}$}  &
\colhead{IR}  &
\colhead{IR Class} &  
\colhead{Notes} \\
\colhead{nr.}    &
\colhead{(J2000)}    &
\colhead{(J2000)}    &
\colhead{(cts)}   &
\colhead{(keV)} &
\colhead{} &
\colhead{} &
\colhead{} &
\colhead{}
}
\startdata
1    & 20 54 38.03 & +77 39 13.03 & 11$\pm$4   & 2.66 & 0.46
& y & ...   & 1,2   \\
2    & 20 54 39.93 & +77 40 00.51 & 9$\pm$4    & 2.40 & 0.40
& y & ...  & 1,2  \\
3                    & 20 54 42.76 & +77 31 49.17 & 7$\pm$3    & 1.81 & 0.83
& n & ... &   \\
4    & 20 54 45.42 & +77 29 35.74 & 13$\pm$4   & 2.70 & 0.38
& n & ... &  1  \\
5                    & 20 54 46.34 & +77 36 58.24 & 12$\pm$4   & 3.65 & 0.47
& y  & ...  & 2,3  \\
6                    & 20 54 51.05 & +77 33 11.31 & 11$\pm$3   & 2.65 & 0.60
& y & agn &   \\
7                    & 20 55 00.39 & +77 40 37.38 & 27$\pm$6   & 3.12 & 0.55
& y & agn &   \\
8                    & 20 55 06.20 & +77 30 03.59 & 57$\pm$8   & 2.38 & 0.96
& y & agn &   \\
9                   & 20 55 14.38 & +77 34 03.69 & 18$\pm$5   & 2.42 & 0.36
& y & agn &   \\
10                  & 20 55 27.94 & +77 35 22.62 & 135$\pm$12 & 1.29 & 0.99
& y & none/star           &  4  \\
11                  & 20 55 34.40 & +77 34 36.88 & 18$\pm$4   & 3.20 & 0.55
& y & ...  & 2  \\
12                  & 20 55 46.39 & +77 29 42.88 & 7$\pm$3    & 3.54 & 0.44
& n & ... &   \\
13                  & 20 55 50.22 & +77 30 32.87 & 20$\pm$5   & 2.54 & 0.48
& n & ... &   \\
14                  & 20 56 00.44 & +77 41 05.33 & 8$\pm$3    & 2.46 & 0.43
& y  & galaxy/yso? & 3,8  \\
15                  & 20 56 03.86 & +77 30 16.28 & 16$\pm$4   & 3.88 & 0.77
& y & ...  &  2  \\
16  & 20 56 13.39 & +77 45 21.47 & 10$\pm$4   & 3.77 & 0.67
& y      & yso (cl. I/II) & 1,3  \\
17                  & 20 56 15.06 & +77 30 40.38 & 7$\pm$3    & 2.08 & 0.47
& y & ...  &  2 \\
18                  & 20 56 15.77 & +77 39 21.97 & 25$\pm$5   & 1.11 & 0.47
& y & none &   \\
19                  & 20 56 16.58 & +77 29 21.25 & 22$\pm$5   & 2.52 & 0.35
& y & ...  & 2  \\
20                  & 20 56 17.15 & +77 40 27.23 & 7$\pm$3    & 2.90 & 0.49
& y & ...  &  2 \\
21                  & 20 56 21.97 & +77 32 37.25 & 13$\pm$4   & 3.76 & 0.57
& y & galaxy/agn &   \\
22  & 20 56 26.19 & +77 28 44.39 & 10$\pm$3   & 2.33 & 0.62
& y & ...  & 1,2   \\
23                  & 20 56 27.47 & +77 42 52.57 & 21$\pm$5   & 3.23 & 0.50
& y & agn &   \\
24                  & 20 56 40.29 & +77 34 20.29 & 27$\pm$5   & 3.92 & 0.83
& y & galaxy/agn &   \\
25                  & 20 57 05.32 & +77 42 25.21 & 22$\pm$5   & 2.71 & 0.50
& y & galaxy &   \\
26                  & 20 57 06.72 & +77 36 56.10 & 5$\pm$2    & 5.10 & 0.34
& y & yso (cl. I)      & HH200-IRS   \\
27                  & 20 57 08.23 & +77 36 37.76 & 10$\pm$3   & 3.57 & 0.33
& y & agn &   \\
28                  & 20 57 09.55 & +77 32 42.98 & 10$\pm$3   & 3.64 & 0.60
& y & none &   \\
29                  & 20 57 15.56 & +77 37 53.63 & 81$\pm$9   & 1.39 & 0.15
& y & none/star & 5  \\
30                  & 20 57 16.88 & +77 36 58.33 & 12$\pm$3   & 2.35 & 0.47
& y & yso (cl. II) &   \\
31                  & 20 57 26.69 & +77 28 50.90 & 57$\pm$8   & 2.63 & 0.99
& y & galaxy/agn &   \\
32                  & 20 57 29.17 & +77 32 45.55 & 6$\pm$2    & 4.48 & 0.33
& y & agn &   \\
33                  & 20 57 32.51 & +77 29 04.68 & 8$\pm$3    & 3.25 & 0.44
& y & ...  & 2  \\
34                  & 20 57 35.86 & +77 29 19.31 & 9$\pm$3    & 2.86 & 0.44
& y & galaxy/agn &   \\
35                  & 20 57 36.97 & +77 29 03.02 & 9$\pm$3    & 1.99 & 0.47
& y & agn &   \\
36                  & 20 57 38.62 & +77 34 12.32 & 25$\pm$5   & 1.40 & 0.47
& y & none/yso         & 6; H$\alpha$ em. star   \\
37                  & 20 57 38.82 & +77 33 34.92 & 40$\pm$6   & 3.58 & 0.40
& y & galaxy/agn &   \\
38                  & 20 57 40.24 & +77 28 38.71 & 123$\pm$12 & 2.38 & 0.59
& y & agn &   \\
39                  & 20 57 44.26 & +77 34 19.56 & 8$\pm$3    & 3.99 & 0.48
& y & agn &   \\
40                  & 20 57 52.30 & +77 37 30.13 & 8$\pm$3    & 3.56 & 0.65
& y & agn &   \\
41                  & 20 57 54.81 & +77 36 09.83 & 14$\pm$4   & 4.05 & 0.39
& y & agn &   \\
42                  & 20 57 55.34 & +77 29 39.00 & 24$\pm$5   & 2.53 & 0.33
& y & galaxy &   \\
43                  & 20 58 04.93 & +77 39 42.30 & 29$\pm$5   & 2.88 & 0.72
& y & agn &   \\
44                  & 20 58 05.77 & +77 36 48.77 & 21$\pm$5   & 2.71 & 0.45
& y & agn &   \\
45  & 20 58 06.67 & +77 44 02.47 & 9$\pm$4    & 1.70 & 0.65
& y & none/star &  1,7 \\
46                  & 20 58 11.03 & +77 33 18.63 & 6$\pm$2    & 3.29 & 0.53
& y & none &   \\
47                  & 20 58 21.78 & +77 36 15.76 & 17$\pm$4   & 3.60 & 0.46
& y & galaxy/agn &   \\
48  & 20 58 25.18 & +77 27 05.70 & 10$\pm$4   & 3.06 & 0.38
& y & agn &  1 \\
49                  & 20 58 37.61 & +77 33 13.36 & 37$\pm$6   & 1.25 & 0.51
& y & none &   \\
50                  & 20 58 38.98 & +77 37 38.94 & 127$\pm$11 & 2.54 & 0.98
& y & agn &   \\
51                  & 20 58 39.79 & +77 31 11.28 & 10$\pm$4   & 1.79 & 0.34
& y & ...  & 2  \\
52                  & 20 58 40.84 & +77 38 20.07 & 11$\pm$3   & 2.40 & 0.55
& n & ... &   \\
53                  & 20 58 46.78 & +77 40 25.47 & 75$\pm$9   & 1.96 & 0.09
& y & yso (cl. II) & 9; H$\alpha$ em. star   \\
54                  & 20 58 57.09 & +77 41 44.97 & 24$\pm$6   & 2.80 & 0.38
& n & ... &   \\
55                  & 20 59 05.10 & +77 34 46.39 & 69$\pm$9   & 2.63 & 0.11
& y & agn &   \\
56                  & 20 59 15.26 & +77 33 22.87 & 30$\pm$6   & 1.32 & 0.62
& y & yso (cl. II) &   \\
57                  & 20 59 15.51 & +77 34 55.20 & 6$\pm$3    & 3.75 & 0.40
& y & ...  &  2  \\
58  & 20 59 18.68 & +77 30 57.72 & 14$\pm$4   & 2.43 & 0.20
& y  & agn & 1,3  \\
59                  & 20 59 19.10 & +77 42 10.71 & 13$\pm$5   & 3.63 & 0.62
& n & ... &   \\
60  & 20 59 42.55 & +77 38 06.20 & 19$\pm$5   & 2.23 & 0.40
& y & none &  1  \\
\enddata
\tablenotetext{a}{X-ray data are from ACIS-I using events in the 0.3 to 7 keV range. 
Tabulated quantities are: running source number; X-ray source position (R.A., Decl.) from {\em wavdetect}
based on CXC events list with no additional astrometric refinement;
net counts and net counts error 
obtained from {\em wavdetect} accumulated in a 36,799 s exposure, rounded to the nearest integer, background subtracted and PSF-corrected;
mean photon energy $\overline{\rm E}$; X-ray variability probability  P$_{var}$ from {\em glvary}. 
A $y$ or $n$ in the IR column denotes whether a potential IR counterpart was found (see Table 4).
IR Class is based on IR color analysis using the IRAC criteria of Gutermuth et
al. (2008) and WISE criteria of Koenig et al. (2012). }
\tablecomments{ \\
1. Source lies near edge of ACIS-I array and is heavily vignetted. \\
2. No IR class assigned due to incomplete IR photometry. \\
3. IR offset $>$2$''$. IR match questionable.  See Table 4 for IR positions. \\
4. Optical counterpart  {\em HST} GSC J205527.93$+$773522.1.
   Classified as  stellar in the GSC. \\
5. Optical counterpart  {\em HST} GSC J205715.70$+$773753.8.
   Classified as  stellar in the GSC. \\
6. Optical counterpart is H$\alpha$ emission star nr. 38 in Ogura \& Sato (1990). \\
7. Star. Tycho J205806.80$+$774402.44. \\
8. IR classificication is ambiguous. \\
9. Optical counterpart is H$\alpha$ emission star nr. 42 in Ogura \& Sato (1990). \\
}
\end{deluxetable}

\clearpage
\begin{deluxetable}{lllll}
\tabletypesize{\scriptsize}
\tablewidth{0pt}
\tablecaption{IR Counterparts to X-ray Sources in  L1228}
\tablehead{
\colhead{CXO nr.}      &
\colhead{2MASS}    &
\colhead{WISE}     &   
\colhead{Spitzer}  &
\colhead{Notes}
}     
\startdata
1  &                    &                         &J205437.98+773911.7 (1.3)  & 3   \\ 
2  &                    &                         &J205440.45+774002.7 (2.7)  & 3,6 \\
3  &                    &                         &                     & 1 \\
4  &                    &                         &                     & 1 \\
5  &                    &                         &J205447.00+773656.6 (2.7) & 2,6 \\
6  &                    &J205451.29+773311.4 (0.8)     &                     & 1  \\
7  &                    &J205500.21+774037.2 (0.6)     &J205500.28+774036.8 (0.7) &    \\
8  &                    &J205505.85+773005.1 (1.9)     &                     & 1 \\
9  &                    &J205514.07+773405.0 (1.7)     &J205514.19+773404.4 (0.9) & 2 \\
10 &J205527.94+773522.3 (0.3) &J205527.89+773522.7 (0.2) &J205527.98+773522.7 (0.2) & 7 \\
11 &                    &                         &J205534.17+773437.9 (1.2) & 7 \\
12 &                    &                         &                     & 2 \\
13 &                    &                         &                     & 2 \\
14 &J205600.25+774105.4 (0.6) &J205559.46+774103.6 (3.6)      &J205559.57+774103.9 (3.1)  & 5,6,7 \\  
15 &                    &                         &J205603.94+773015.8  (0.6) & 4 \\
16 &                    &                         &J205612.79+774518.4  (3.6) & 6 \\
17 &                    &                         &J205614.93+773040.8  (0.6) &   \\
18 &J205615.81+773922.5 (0.5) &J205615.73+773922.2 (0.3)      &J205615.94+773921.9 (0.5)  &   \\
19 &                    &                         &J205616.45+772921.3 (0.4) & 4 \\
20 &                    &                         &J205617.21+774028.1 (0.9) & \\
21 &                    &J205621.51+773237.5 (1.5)     &J205622.03+773237.6 (0.4)  & 7 \\
22 &                    &                         &J205625.85+772843.4 (1.5) & 4 \\
23 &                    &J205627.39+774253.1 (0.6)     &J205627.29+774253.0 (0.7)  & \\
24 &                    &J205640.35+773421.1 (0.8)     &J205640.29+773420.1 (0.2)  & \\
25 &                    &                         &J205705.49+774226.6 (1.5)  & 7 \\
26 &J205706.55+773655.9 (0.6) &J205707.16+773657.4 (1.9)      &J205706.61+773656.0 (0.4)  & 5,8,9 \\                 
27 &                    &J205708.44+773637.7 (0.7)     &J205708.23+773637.8 (0.0) & 9 \\
28 &                    &                         &J205709.66+773242.8 (0.4)  & \\
29 &J205715.60+773754.0 (0.4) &J205715.63+773753.8 (0.3)      &J205715.53+773753.8 (0.2)  & \\
30 &J205716.97+773658.6 (0.4) &J205716.98+773658.5 (0.4)      &J205717.02+773658.4 (0.5)  & 5,9,10 \\
31 &                    &J205726.65+772851.4 (0.5)     &J205726.58+772851.2 (0.5)  & \\
32 &                    &                         &J205729.15+773245.4 (0.2) & \\
33 &                    &                         &J205732.72+772904.2 (0.8) & \\
34 &                    &J205735.43+772918.5 (1.6)     &J205735.79+772919.4 (0.2)  & 7 \\
35 &                    &J205737.40+772902.3 (1.6)     &J205737.15+772902.9 (0.6)  & \\
36 &J205738.73+773412.5 (0.4) &J205738.71+773412.4 (0.3)      &J205738.73+773412.3 (0.4)  & 11 \\
37 &                    &J205738.82+773335.1 (0.2)     &J205738.79+773334.8 (0.2)  & \\
38 &                    &J205740.44+772838.8 (0.7)     &J205740.46+772838.9 (0.7)  & \\
39 &                    &                         &J205744.35+773419.5 (0.3) & \\
40 &                    &                         &J205752.26+773730.2 (0.2)  & \\
41 &                    &                         &J205754.83+773610.3 (0.5)  & \\
42 &                    &                         &J205755.09+772938.1 (1.2) & 7 \\
43 &                    &                         &J205804.99+773942.8 (0.5)  & \\
44 &                    &J205805.86+773648.5 (0.4)     &J205805.86+773648.8 (0.4)  & \\
45 &J205806.81+774402.4 (0.5) &J205806.82+774402.3 (0.5)      &J205806.83+774402.5 (0.5) & 3 \\
46 &                    &                         &J205811.25+773318.8 (0.7) & \\
47 &                    &J205821.91+773616.0 (0.5)      &J205821.70+773616.0 (0.4)  & \\
48 &                    &J205825.16+772706.3 (0.6)      &J205825.28+772705.7 (0.3)  & \\
49 &J205837.68+773314.0 (0.7) &J205837.72+773313.7 (0.5)      &J205837.62+773314.0 (0.6)  & \\
50 &                    &J205839.07+773739.0 (0.3)     &J205839.00+773738.9 (0.1)  & \\
51 &                    &                         &J205839.96+773111.4 (0.6) & \\
52 &                    &                         &                     & 2 \\
53 &J205846.68+774025.6 (0.3) &J205846.71+774025.6 (0.3)      &J205846.71+774026.0 (0.6)  & 2,9,12 \\
54 &                    &                         &                     & 1 \\
55 &                    &                         &J205905.14+773447.5 (1.1) & \\
56 &J205915.41+773322.8 (0.5) &J205915.40+773322.8 (0.5)     &J205915.37+773322.7 (0.4)  & \\
57 &                    &                         &J205915.42+773455.2 (0.3) & 4,7 \\
58 &                    &J205919.39+773056.7 (2.5)     &J205919.28+773056.7 (2.2)  & 6 \\
59 &                    &                         &                     & 1 \\
60 &J205942.32+773808.3 (2.2) &J205942.35+773808.1 (2.0)      &                     & 1,6 \\
\enddata
\tablecomments{   
The number in parentheses  is the offset in arcseconds
between the IR position and the X-ray position in Table 3. \\  
1. Source lies outside {\em Spitzer} coverage area. \\
2. Source is off-edge in IRAC 1-4. \\
3. Source is off-edge in IRAC 1 \& 3. \\
4. Source is off-edge in IRAC 2 \& 4. \\
5. Source appears extended in IRAC images. \\
6. IR position is offset by $>2$$''$ from X-ray position. Match is questionable. \\
7. Multiple IR sources near X-ray  position. Possible source confusion. \\
8. HH 200-IRS. Previously identified in Table 1 of Devine et al. (2009) and as
   source 5 in Table 4 of Chapman and Mundy (2009). 
   The radio counterpart is L1228 VLA 4  in Table 2 of Reipurth et al. (2004). 
   Source is extended in 2MASS, IRAC, and WISE images. Close IR companion (Fig. 6). \\
9. WISE emission is likely variable ($varflg \geq$ 6 in one or more bands). \\
10. Corresponds  to source 9 in Table 4 of Chapman and Mundy (2009) and source 11 of Kirk et al. (2009). \\
11. Optical counterpart is source 38 (OSHA 38) of Ogura \& Sato (1990). \\
12. Optical counterpart is source OSHA 42 in Table 1 of Kun et al. (2009)
    and source 42 of Ogura \& Sato (1990). \\
}
\end{deluxetable}

\clearpage
\begin{deluxetable}{llllrrrrrllll}
\rotate
\tabletypesize{\scriptsize}
\tablewidth{0pt}
\tablecaption{L1228 IR Photometry\tablenotemark{a}}
\tablehead{
\colhead{CXO nr.}      &
\colhead{J}    &
\colhead{H}     &
\colhead{K$_{s}$}     &
\colhead{I1}      &
\colhead{I2}    &
\colhead{I3}     &
\colhead{I4}     &
\colhead{M1}     &
\colhead{W1}      &
\colhead{W2}    &
\colhead{W3}     &
\colhead{W4} 
}     
\startdata
1  &             &              &              & ...    ...   & 17.58  0.27 & ...    ...  & ...    ...  & ...    ...  &              &             &             &               \\
2  &             &              &              & ...    ...   & 17.31  0.23 & ...    ...  & 16.16  0.27 & ...    ...  &              &             &             &               \\
3  &             &              &              & ...    ...   & ...    ...  & ...    ...  & ...    ...  & ...    ...  &              &             &             &               \\
4  &             &              &              & ...    ...   & ...    ...  & ...    ...  & ...    ...  & ...    ...  &              &             &             &               \\
5  &             &              &              & ...    ...   & ...    ...  & ...    ...  & ...    ...  &  9.90  0.19 &              &             &             &               \\
6  &             &              &              & ...    ...   & ...    ...  & ...    ...  & ...    ...  &  ...   ...  & 17.73  0.18  & 16.51  0.22 & 13.27  ...  &   9.57  ...   \\ 
7  &             &              &              & 15.98  0.06  & 15.73  0.12 & 15.17  0.09 & 14.40  0.11 & 10.89  0.13 & 16.51  0.07  & 15.88  0.12 & 13.12  ...  &   9.56  ...   \\
8  &             &              &              & ...    ...   & ...    ...  & ...    ...  & ...    ...  & ...    ...  & 16.31  0.06  & 15.42  0.08 & 13.19  ...  &   9.35  ...   \\
9  &             &              &              & ...     ...  & ...    ...  & ...    ...  & ...    ...  & 9.13   0.15 & 15.57  0.04  & 14.91  0.06 & 12.24  0.23 &   9.12  0.35  \\
10 & 13.88  0.03 & 13.26   0.03 & 13.07   0.05 & 12.78  0.05  & 12.76  0.06 & 12.70  0.05 & 12.64  0.05 & ...    ...  & 12.85  0.02  & 12.65  0.02 & 12.27  0.24 &   9.30  ...   \\
11 &             &              &              & ...    ...   & 16.71  0.18 & ...    ...  & 15.22  0.18 & ...    ...  &              &             &             &               \\
12 &             &              &              & ...    ...   & ...    ...  & ...    ...  & ...    ...  &             &              &             &             &               \\
13 &             &              &              & ...    ...   & ...    ...  & ...    ...  & ...    ...  & $>$10.80    &              &             &             &               \\
14 &             &              & $>$13.93     & 13.39  0.06  & 13.28  0.07 & 11.85  0.06 &  9.90  0.06 &             & 13.12  0.02  & 12.85  0.02 &  8.80  0.02 &   6.73  0.06  \\
15 &             &              &              & 16.44  0.06  & ...    ...  & 15.67  0.10 & ...    ...  &             &              &             &             &               \\
16 &             &              &              & 17.96  0.28  & 17.03  0.20 & 16.30  0.36 & 16.56  0.30 & $>$10.99    &              &             &             &               \\
17 &             &              &              & 18.51  0.07  & ...    ...  & ...    ...  & ...    ...  & $>$11.86    &              &             &             &               \\
18 & 10.39  0.03 &  9.98  0.03  &  9.98  0.03  &  9.84  0.05  &  9.84  0.05 &  9.82  0.05 &   9.78 0.05 &  9.77  0.11 &  9.79  0.02  & 9.84  0.02  &  9.87 0.03  &  9.27   ...  \\
19 &             &              &              & 17.88  0.06  & ...    ...  & 16.37  0.21 & ...    ...  &             &              &             &             &               \\
20 &             &              &              & 17.45  0.22  & 17.03  0.20 & ...    ...  & 15.95  0.21 &             &              &             &             &               \\
21 &             &              &              & 17.59  0.06  & 17.26  0.23 & 17.04  0.28 & 16.61  0.33 & ...    ...  &  17.54  0.15 & 17.13 ...   & 13.25 ...   &  9.65  ...    \\
22 &             &              &              & $>$17.85     & ...    ...  & ...    ...  & ...    ...  & ...    ...  &              &             &             &               \\
23 &             &              &              & 15.09  0.05  & 14.01  0.07 & 12.92  0.06 & 11.77  0.06 &  8.48  0.11 &  15.71  0.04 & 14.00 0.04  & 10.91  0.07 &  8.58  0.22   \\
24 &             &              &              & 17.17  0.05  & 16.71  0.18 & 16.21  0.06 & 15.88  0.24 &             &  16.74  0.08 & 16.07 0.14  & 12.49  0.24 &  9.62  0.45   \\
25 &             &              &              & 16.71  0.06  & 16.20  0.05 & 16.03  0.15 & 14.76  0.06 & 10.92  0.13 &              &             &             &               \\
26\tablenotemark{b,c} & 14.76  0.09 & 13.56  0.08  & 12.71  0.05  & 11.37  0.06  & 10.57  0.06 &  9.76  0.06 &  8.85  0.05 &  3.38  0.11 &  11.75  0.02 & 10.63 0.02  &  7.42  0.01 &  3.43  0.01   \\
27\tablenotemark{c} &             &              &              & 16.45  0.13  & 15.17  0.10 & 13.87  0.08 & 13.32  0.08 & ...    ...  &  16.84  0.09 & 14.69 0.05  & 12.26  0.20 &  7.58  0.08   \\
28 &             &              &              & 18.16  0.06  & 17.30  0.06 & 17.42  0.13 & 17.18  0.26 &             &              &             &             &               \\
29 & 12.14  0.02 & 11.48  0.03  & 11.10  0.02  & 10.66  0.05  & 10.60  0.06 & 10.51  0.06 & 10.46  0.05 &             & 10.86  0.02  & 10.56 0.02  & 10.24  0.04 &  8.74  0.22   \\
30\tablenotemark{c} & 13.51  0.03 & 11.98  0.03  & 11.00  0.02  &  9.96  0.05  &  9.24  0.05 &  8.76  0.05 &  8.04  0.05 &  4.91  0.11 & 10.03  0.02  &  9.14 0.02  &  7.09  0.01 &  4.72  0.02   \\
31 &             &              &              & 16.47  0.05  & 16.04  0.06 & 15.57  0.06 & 15.26  0.12 &             & 16.91  0.09  & 16.43 0.19  & 12.15  0.18 &  9.74  0.53   \\
32 &             &              &              & 16.21  0.05  & 15.62  0.05 & 14.92  0.11 & 13.79  0.06 &             &              &             &             &               \\
33 &             &              &              & 17.59  0.06  & 16.65  0.07 & 16.20  0.07 & $>$15.39    &             &              &             &             &               \\
34 &             &              &              & 16.65  0.05  & 16.33  0.06 & 16.00  0.06 & 14.71  0.16 &             & 17.12  0.11  & 16.56 0.21  & 12.26  0.20 &  9.11  ...    \\
35 &             &              &              & 16.77  0.05  & 15.74  0.12 & 14.41  0.05 & 13.61  0.09 &  9.83  0.11 & 17.73  0.18  & 15.78 0.11  & 11.40  0.09 &  8.92  0.25   \\
36 & 11.98  0.02 & 11.05  0.03  & 10.73  0.02  & 10.60  0.05  & 10.54  0.06 & 10.47  0.06 & 10.46  0.06 & 10.60  0.11 & 10.62  0.02  & 10.49 0.02  & 10.34  0.04 &  8.91  0.25   \\
37 &             &              &              & 16.65  0.05  & 16.06  0.05 & 15.56  0.06 & 14.89  0.07 & 10.87  0.11 & 16.68  0.08  & 16.37 0.18  & 12.44  0.22 &  9.47  ...    \\
38 &             &              &              & 16.38  0.05  & 15.65  0.12 & 14.64  0.06 & 13.72  0.10 & 10.50  0.12 & 16.74  0.08  & 15.54 0.10  & 12.04  0.16 &  9.43  0.41   \\
39 &             &              &              & 17.36  0.06  & 16.42  0.06 & 15.75  0.06 & 14.21  0.06 & ...    ...  &              &             &             &               \\
40 &             &              &              & 18.20  0.06  & 17.54  0.06 & 16.56  0.08 & 16.00  0.09 & 12.03  0.18 &              &             &             &               \\
41 &             &              &              & 17.38  0.05  & 16.59  0.06 & 16.23  0.19 & 15.24  0.07 & 11.52  0.14 &              &             &             &               \\
42 &             &              &              & 16.78  0.15  & 16.25  0.16 & 15.86  0.16 & 14.38  0.12 & 10.40  0.11 &              &             &             &               \\
43 &             &              &              & 17.02  0.05  & 16.45  0.06 & 16.08  0.06 & 15.83  0.22 & $>$10.77    &              &             &             &               \\
44 &             &              &              & 16.99  0.05  & 16.19  0.05 & 15.47  0.06 & 14.26  0.06 & 10.54  0.11 & 17.09  0.10  & 15.70 0.10  & 11.70  0.12 &  9.35  ...    \\
45 & 10.79  0.02 & 10.50  0.03  & 10.47  0.02  & ...    ...   & 10.48  0.06 & ...    ...  & 10.43  0.06 &             & 10.41  0.02  & 10.43 0.02  & 10.37  0.04 &  9.40  ...    \\
46 &             &              &              & 16.39  0.13  & 16.01  0.13 & 15.76  0.15 & 15.68  0.20 &             &              &             &             &               \\
47 &             &              &              & 16.75  0.05  & 15.78  0.12 & 14.89  0.11 & 13.88  0.09 & 10.72  0.12 & 16.90  0.09  & 16.01 0.13  & 12.00  0.16 &  9.53  0.45   \\
48 &             &              &              & 16.34  0.05  & 16.07  0.06 & 15.71  0.06 & 14.88  0.18 &             & 16.71  0.08  & 17.16 0.35  & 12.57  0.25 &  9.68  ...    \\
49 & 13.71  0.03 & 13.13  0.04  & 12.84  0.03  & 12.60  0.05  & 12.52  0.05 & 12.42  0.05 & 12.43  0.05 & 11.65  0.14 & 12.69  0.03  & 12.53 0.02  & 11.58  0.10 &  9.16  ...    \\
50 &             &              &              & 16.29  0.05  & 15.40  0.05 & 14.52  0.06 & 13.44  0.06 &             & 16.58  0.07  & 15.12 0.07  & 11.32  0.09 &  8.99  0.28   \\
51 &             &              &              & 18.24  0.06  & ...    ...  & 16.95  0.08 & ...    ...  & $>$12.45    &              &             &             &               \\
52 &             &              &              & ...    ...   & ...    ...  & ...    ...  & ...    ...  & $>$13.87    &              &             &             &               \\
53\tablenotemark{c} & 11.51  0.02 & 10.36  0.03  &  9.70  0.02  & ...    ...   & ...    ...  & ...    ...  & ...    ...  &  4.52  0.11 &  8.75  0.02  &  8.09 0.02  &  6.50  0.01 &  4.45  0.02   \\
54 &             &              &              & ...    ...   & ...    ...  & ...    ...  & ...    ...  & ...    ...  &              &             &             &               \\
55 &             &              &              & 17.65  0.06  & 16.96  0.09 & 16.40  0.08 & 15.06  0.23 & ...    ...  &              &             &             &               \\
56 & 11.90  0.02 & 10.95  0.03  & 10.55  0.02  & 10.25  0.05  &  9.92  0.05 &  9.60  0.05 &  8.84  0.05 &  6.35  0.11 & 10.29  0.02  &  9.95 0.02  &  7.95  0.02 &  6.20  0.04   \\
57 &             &              &              & 18.21  0.29  & ...    ...  & ...    ...  & ...    ...  &             &              &             &             &               \\
58 &             &              &              & 15.83  0.11  & 15.70  0.12 & 15.20  0.12 & 14.44  0.14 & ...    ...  & 15.84  0.05  & 15.33 0.08  & 11.87  0.14 &  9.09  ...    \\
59 &             &              &              & ...    ...   & ...    ...  & ...    ...  & ...    ...  & ...    ...  &              &             &             &               \\
60 & 14.56 0.04  &  13.99 0.04  & 13.59  0.05  & ...    ...   & ...    ...  & ...    ...  & ...    ...  & ...    ...  & 13.49  0.02  & 13.32 0.03  & 11.43  0.09 &  9.72  0.53   \\
199\tablenotemark{d} 
   & 13.02  0.03 & 10.61  0.03  &  9.17  0.03  & ...    ...   & ...    ...  & ...    ...  & ...    ...  & ...    ...  &  7.89  0.02  &  6.55 0.02  &  3.88  0.01 &   1.38  0.01  \\    

\enddata
\tablenotetext{a}{J,H,K$_{s}$ = 2MASS. I1 - I4 = IRAC bands 1-4. M1 = MIPS-24. W1 - W4 = WISE bands 1-4. Each column gives
                  source magnitude followed by its uncertainty.} 
\tablenotetext{b}{There is disagreement between 2MASS JHK$_{s}$ photometry and that obtained with the KPNO 4m
                  by Chapman \& Mundy (2009).}
\tablenotetext{c}{IR emission may be variable. WISE $varflg$ $\geq$ 6 in one or more bands.}
\tablenotetext{d}{HH 199-IRS. Not detected by Chandra.} \\
\end{deluxetable}

\clearpage
\begin{deluxetable}{lllllllc}
\tabletypesize{\scriptsize}
\tablewidth{0pt}
\tablecaption{Spectral Fits of  Lynds 1228 Bright X-ray Sources\tablenotemark{1}}
\tablehead{
\colhead{CXO}      &
\colhead{Model}     &
\colhead{N$_{\rm H}$}    &
\colhead{kT$_{1}$}   &
\colhead{$\Gamma$}     &
\colhead{norm$_{1}$}  &
\colhead{$\chi^2$/dof}  &
\colhead{F$_{x}$(0.3 - 7 keV)}  \\
\colhead{nr.}    &
\colhead{}    &
\colhead{(10$^{22}$ cm$^{-2}$)}    &
\colhead{(keV)}   &
\colhead{} &
\colhead{(10$^{-5}$)} &
\colhead{} &
\colhead{(10$^{-14}$ erg cm$^{-2}$ s$^{-1}$)}
}
\startdata
8  &  1T                  &   1.10 [0.25 - 3.04] &  2.20 [0.83 - 9.88] & ...                 &  3.11 [1.52 - 11.9] & 7.39/8 &    1.94 (4.86) \nl
8  &  PL                  &   1.05 [0.24 - 3.14] &  ...                & 2.55 [1.49 - 4.35]  &  1.63 [0.00 - 16.3] & 8.01/8 &    1.92 (7.59) \nl
10 &  2T\tablenotemark{a} &   0.16 [0.00 - 0.50] &  0.22 [0.14 - 0.51] & ...                 &  1.97 [0.23 -  ...] & 7.61/7 &    3.01 (6.57) \nl
29 &  1T                  &   0.69 [0.35 - 0.99] &  0.84 [0.56 - 1.05] & ...                 &  1.92 [0.91 - 4.14] & 2.99/3 &    1.20 (6.17) \nl
31 &  1T                  &   0.70 [0.06 - 1.78] &  4.09 [1.54 - ...]  & ...                 &  2.10 [1.21 - 4.28] & 2.82/4 &    2.13 (3.56) \nl
31 &  PL                  &   0.66 [0.04 - 1.71] &  ...                & 1.95 [1.02 - 3.12]  &  0.77 [0.00 - 3.03] & 2.92/4 &    2.01 (3.95) \nl
38 &  1T                  &   0.32 [0.09 - 0.65] &  6.38 [3.20 - ...]  & ...                 &  3.26 [2.55 - 4.31] & 11.2/11&    4.35 (5.82) \nl
38 &  PL                  &   0.42 [0.11 - 0.84] &  ...                & 1.83 [1.23 - 2.58]  &  1.23 [0.60 - 2.69] & 11.0/11&    4.06 (6.69) \nl
50\tablenotemark{b} &  1T &   0.31 [0.12 - 0.69] &  20.4 [4.35 - ...]  & ...                 &  3.80 [2.96 - 5.14] & 3.32/6 &    4.96 (6.25) \nl
50\tablenotemark{b} &  PL &   0.35 [0.07 - 0.75] &  ...                & 1.44 [0.97 - 1.99]  &  0.92 [0.49 - 1.78] & 3.31/6 &    4.84 (6.47) \nl
53 &  1T                  &   0.20 [0.01 - 1.45] &  2.71 [0.49 - 4.79] & ...                 &  1.73 [1.21 - 12.4] & 6.62/8 &    1.99 (2.76)\tablenotemark{c}  \nl
53 &  1T                  &   \{0.65\}\tablenotemark{d}  &  1.60 [1.22 - 2.19] & ...         &  2.22 [1.63 - 2.85] & 8.58/9 &    1.59 (3.90)  \nl
55 &  1T                  &   1.84 [0.40 - 3.79] &  2.91 [1.50 - ...]  & ...                 &  4.25 [1.78 - 10.8] & 4.04/4 &    2.68 (6.84) \nl
55 &  PL                  &   1.84 [0.26 - 5.65] &  ...                & 2.23 [0.98 - 4.54]  &  1.87 [0.00 - 53.7] & 5.09/4 &    2.56 (8.81) \nl
\enddata
 \tablenotetext{1}{Based on XSPEC (vers. 12.7.1) fits of background-subtracted ACIS-I spectra binned to
a minimum of 10 counts per bin, unless otherwise noted. 
The tabulated parameters are the CXO source number (Table 3); model type, where PL denotes  a power-law model and 1T or 2T denotes 
the number  of temperature components in the $apec$ optically thin thermal plasma model;
absorption column density (N$_{\rm H}$);
plasma energy (kT$_{1}$);  photon power-law index ($\Gamma$) where F$_{x}$ $\propto$ $E^{-(\Gamma - 1)}$; 
XSPEC component normalization (norm$_{1}$); $\chi^2$ statistic/degrees-of-freedom;
and the absorbed X-ray flux followed in parentheses by the unabsorbed flux. 
For those sources classified as galaxies or AGNs on the basis of IR colors, results of a PL model are given in addition
to thermal models.
Solar abundances were assumed for $apec$ models and are referenced to the values of  Anders \& Grevesse (1989).
Square brackets enclose 90\% confidence intervals and an ellipsis means that
the algorithm used to compute confidence intervals did not converge.
}
\tablenotetext{a}{A 1T model fit is statistically unacceptable. The 2T model requires an additional hot
plasma component at kT$_{2}$ = 1.43 [1.22 - 1.95] keV and  norm$_{2}$ = 1.33 [0.85 - 1.94].
The tabulated results are for the 2T model.}

\tablenotetext{b}{The spectrum was binned to a minimum of 15 counts per bin. A 1T model requires a 
high but uncertain plasma temperature.}

\tablenotetext{c}{Class II YSO (OSHA 42). At an assumed L1228 distance of 200 pc the unabsorbed flux
gives an X-ray luminosity  log L$_{x}$(0.3 - 7 keV) = 29.12 ergs s$^{-1}$.}

\tablenotetext{d}{N$_{\rm H}$ held fixed at the value corresponding to A$_{\rm V}$ = 3.42 mag 
obtained by Kun et al. (2009). The unabsorbed flux gives 
log L$_{x}$(0.3 - 7 keV) = 29.27 ergs s$^{-1}$ at d = 200 pc.}
\end{deluxetable}

\clearpage
\begin{deluxetable}{llllllllll}
\tabletypesize{\scriptsize}
\tablewidth{0pt}
\tablecaption{IR SED Model Fits of  Lynds 1228 YSOs\tablenotemark{1}}
\tablehead{
\colhead{CXO}      &
\colhead{ A$_{\rm V,ism}$}     &
\colhead{ A$_{\rm V,cs}$}    &
\colhead{M$_{disk}$}   &
\colhead{$\dot{\rm M}_{acc}$}  &
\colhead{dist.}  &
\colhead{L$_{tot}$}   &
\colhead{$\alpha$}  &
\colhead{IR class}  &
\colhead{IR class} \\
\colhead{nr.}    &
\colhead{(mag)}    &
\colhead{(mag)}    &
\colhead{(M$_{\odot}$)} &
\colhead{(M$_{\odot}$ yr$^{-1}$)} &
\colhead{(pc)} &
\colhead{(L$_{\odot}$)} &
\colhead{} &
\colhead{($\alpha$)} &
\colhead{(colors)} 
}
\startdata
14\tablenotemark{a}  & ...   & ...                        &  ...            & ...                        & ... & ...     & $+$1.38  & I       & I    \nl
16\tablenotemark{a}  & ...   & ...                        &  ...            & ...                        & ... & ...     & $+$0.31  & I/flat  & I/II \nl
26\tablenotemark{b}  &  5.0  & $>$32\tablenotemark{b}     &  1.1e-04        &  2.0e-06\tablenotemark{c}  & 191 & 0.66    & $-$0.04  & flat    & I    \nl
30                   &  8.9  &$<$0.35\tablenotemark{d}    &  1.3e-03        &  6.5e-07\tablenotemark{c}  & 209 & 0.24    &$-$0.56   & II      & II   \nl
53                   &  3.4  & 2.2                        &  7.9e-03        &  3.8e-07                   & 191 & 0.52    &$-$0.98   & II      & II   \nl
56                   &  1.9  & ...\tablenotemark{c}       &  1.3e-04        &  3.5e-11\tablenotemark{c}  & 200 & 0.23    &$-$1.44   & II      & II   
\enddata
 \tablenotetext{1}{Based on  fits of JHK$_{s}$ and {\em Spitzer} IRAC/MIPS-24 and WISE IR photometry using the 
Robitaille et al. (2007) modeling tool. 
The source number in the first column refers to Table 3. Tabulated parameters determined from the modeling
tool are the interstellar and circumstellar extinctions. disk mass, mass accretion rates (sum of disk and
envelope contributions),
best-fit distance (constrained during fitting to lie within the range 180 - 220 pc), and total system 
luminosity L$_{bol}$. The quoted values are the median values of the five fitted models with the 
lowest $\chi^2$ fit statistic, unless otherwise noted.
The IR spectral index $\alpha$ is based on power-law fits over the $\approx$2 - 8 $\mu$m range where
$\alpha$ = $d$~log($\lambda$F$_{\lambda}$)/$d$~log$\lambda$. 
The IR class is  based on the spectral index 
criteria of Haisch et al. (2001) and that   determined
from mid-IR colors (see Fig. 5).
} 
\tablenotetext{a}{No SED modeling performed due to lack of near-IR photometry.}
\tablenotetext{b}{HH 200-IRS. The fit used more recent JHK$_{s}$ photometry from Chapman \& Mundy (2009) 
                  instead of 2MASS data. The  A$_{\rm V,cs}$ limit is the lowest value obtained from 
                  the set of best-fit models.}
\tablenotetext{c}{Parameter is not tightly constrained by IR  SED models. Fitted values 
                  for different models span an order of magnitude.}
\tablenotetext{d}{The  A$_{\rm V,cs}$ limit is the largest  value obtained from the set of best-fit models.}

\end{deluxetable}

\clearpage
\begin{figure}
\figurenum{1}
\epsscale{1.0}
\includegraphics*[width=15.0cm,height=10.845cm,angle=0]{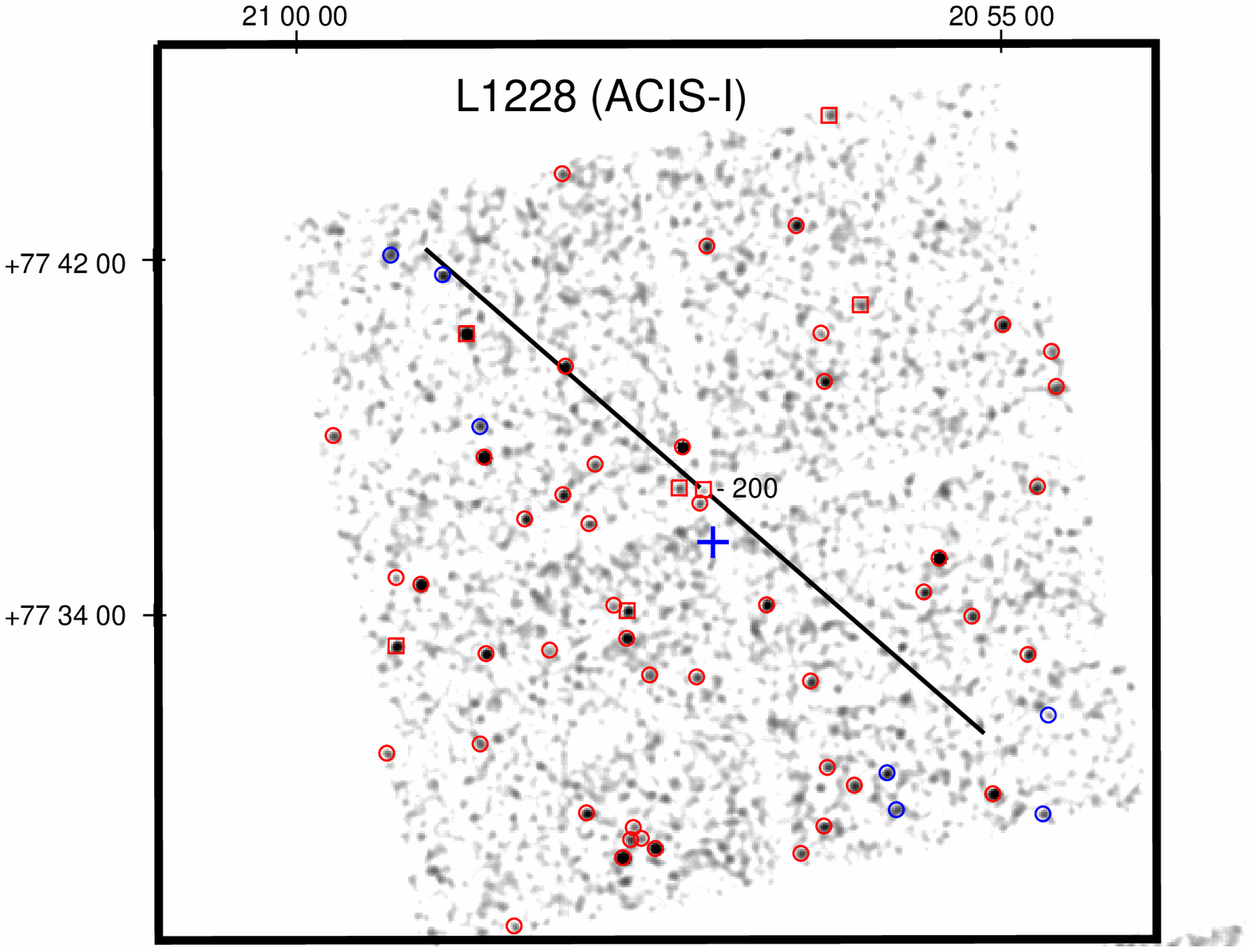}
\caption{Positions of L1228 X-ray sources overlaid on the {\em Chandra} ACIS-I image (0.3 - 7 keV). 
 The image has been Gaussian-smoothed with a 3-pixel kernel. The raw (unsmoothed) pixel
 size is 0.$''$49.   Objects marked in red have
 candidate IR counterparts (Table 4) and those in blue  lack counterparts.
 Squares denote known or suspected YSOs (Table 3). The object labeled  ``200'' is
 HH 200-IRS (CXO source nr. 26 in Table 3). The solid lines show the HH 200 bipolar
 outflow axes at P.A. = 49$^{\circ}$/229$^{\circ}$, measured east from north (Devine et al. 2009).
 A cross ($+$) marks the nominal pointing position.  Log intensity scale. 
 J2000 coordinate overlay.}
\end{figure}

\clearpage
\begin{figure}
\figurenum{2}
\epsscale{1.0}
\includegraphics*[width=14.0cm,height=18.12cm,angle=0]{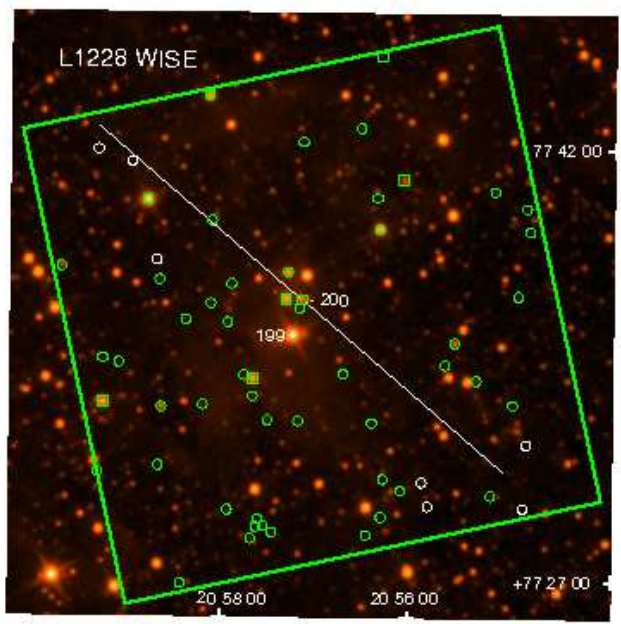}
\caption{Positions of L1228 X-ray sources overlaid on a WISE W1 (3.4 $\mu$m) image.
Squares denote known or suspected YSOs. Objects marked in green have IR counterparts
but only the brightest IR sources are visible in this image.
Objects marked in white circles lack
IR counterparts. The HH-driving sources HH 199-IRS (IRAS 20582$+$7724) and HH 200-IRS are marked.
The solid lines show the HH 200 outflow axes at P.A. = 49$^{\circ}$/229$^{\circ}$.
The emission knots associated with HH 199 are directed generally eastward (to the left) but
span a range of position angles P.A. = 60$^{\circ}$ - 100$^{\circ}$ (Devine et al. 2009; see also Fig. 7). 
HH 199-IRS  was not detected in X-rays.  The {\em Chandra} ACIS-I detector footprint
is shown as a 16.9$'$ $\times$ 16.9$'$ square. Log intensity scale. J2000 coordinates.} 
\end{figure}

\clearpage
\begin{figure}
\figurenum{3}
\epsscale{1.0}
\includegraphics*[width=11.1cm,height=15.0cm,angle=-90]{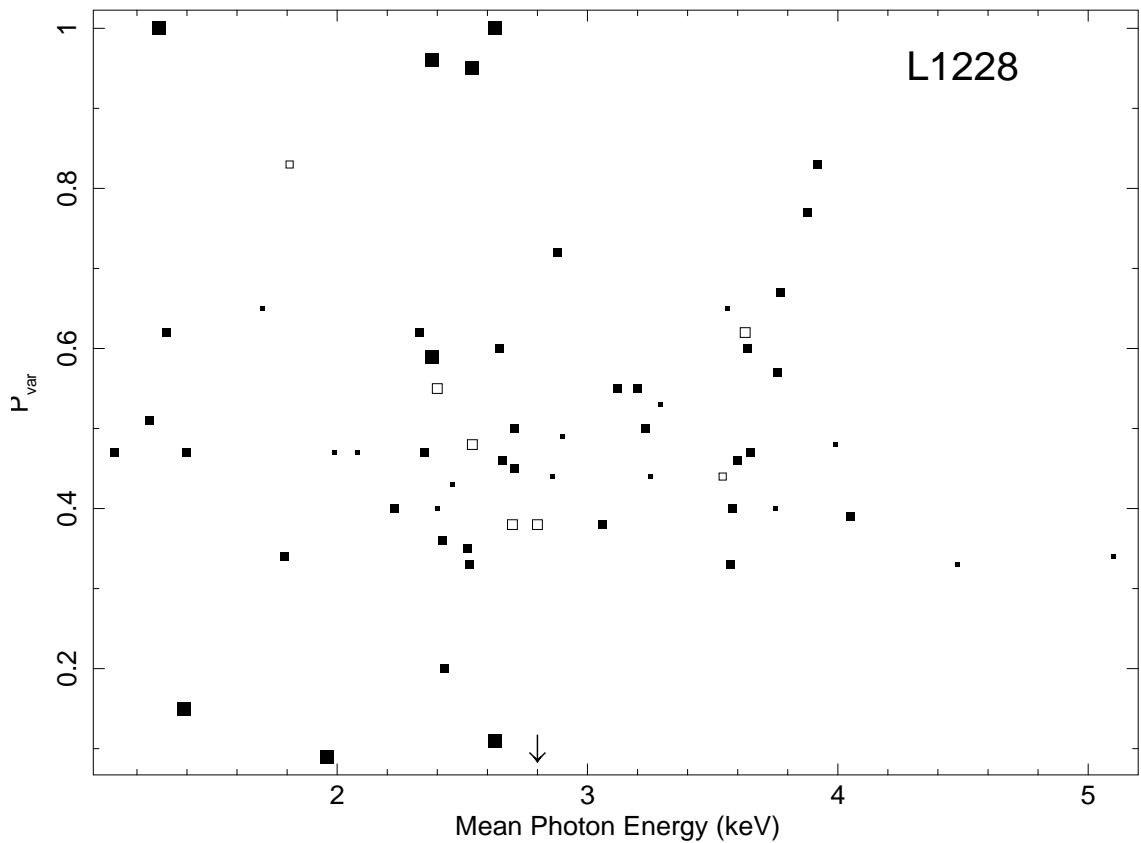}
\caption{X-ray variability probability P$_{var}$ versus mean photon energy for the X-ray
 sources detected in L1228. Symbol sizes reflect the number of X-ray counts:
 small $<$10  cts, medium 10 - 50 cts, large $>$50 cts. Filled symbols denote
 sources with IR counterparts and open symbols lack counterparts. The arrow
 at 2.8 keV marks the average value of the mean photon energy for the X-ray sample. 
 The source with the highest mean energy at far right is HH 200-IRS. Those
 sources with P$_{var}$ $\geq$ 0.9 all have $>$50 counts (CXO sources 8, 10, 31, 50). }
\end{figure}

\clearpage
\begin{figure}
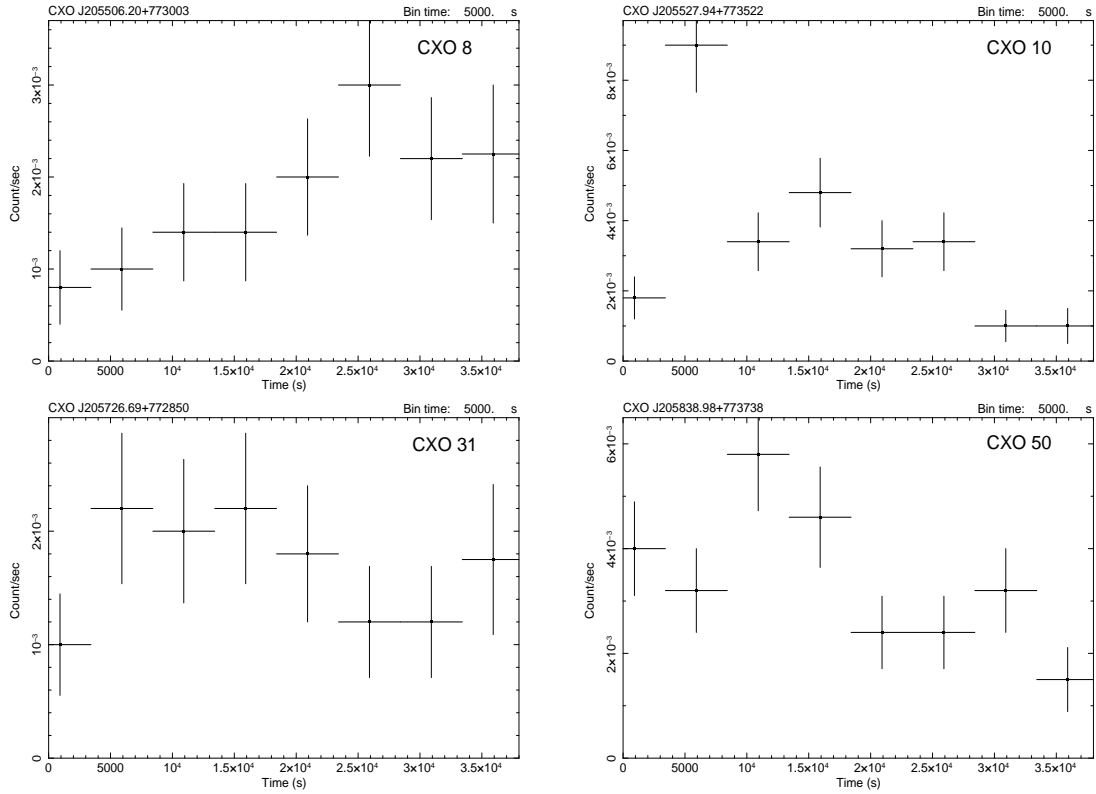

\figurenum{4}
\epsscale{1.0}
\includegraphics*[width=5.25cm,height=7.5cm,angle=-90]{f4a.eps}
\includegraphics*[width=5.25cm,height=7.5cm,angle=-90]{f4b.eps} \\
\includegraphics*[width=5.25cm,height=7.5cm,angle=-90]{f4c.eps}
\includegraphics*[width=5.25cm,height=7.5cm,angle=-90]{f4d.eps}

\caption{Chandra ACIS-I light curves (0.3 - 7 keV) of L1228  X-ray sources 
with high probability of variability P$_{var}$ $>$ 0.9, binned at 5000 s intervals.}
\end{figure}

\clearpage
\begin{figure}
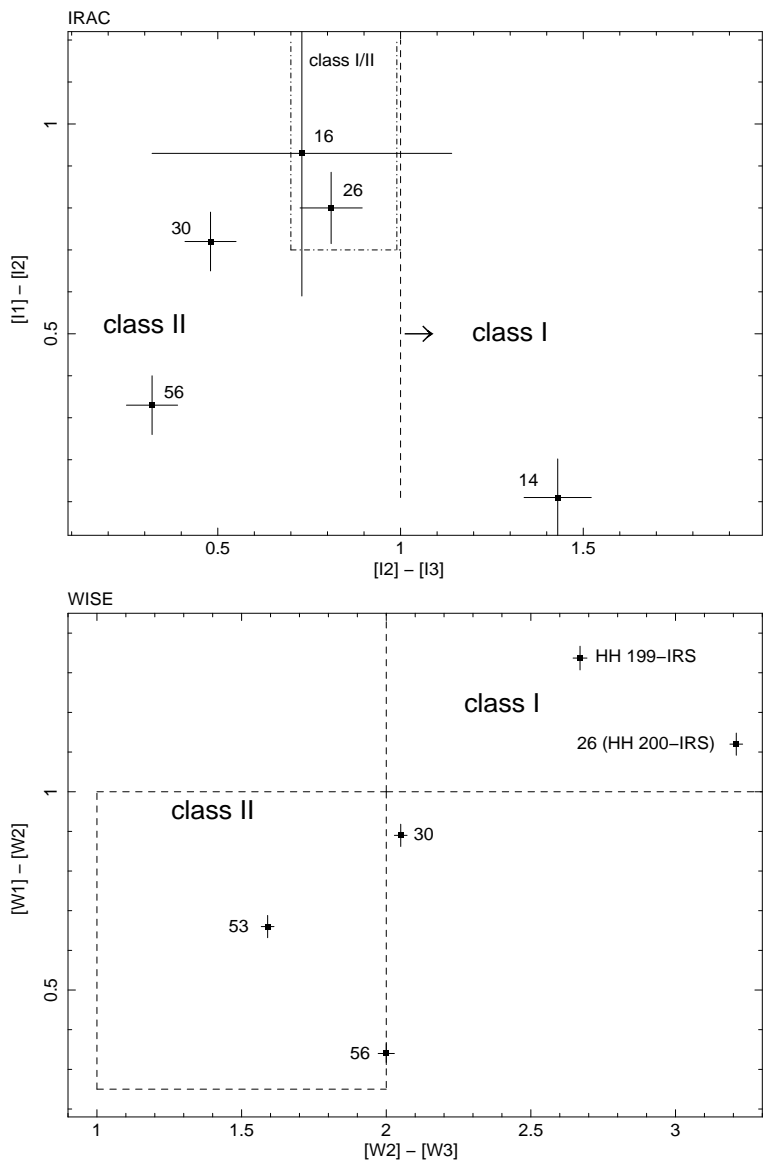

\figurenum{5}
\epsscale{1.0}
\includegraphics*[width=7.7cm,height=10.0cm,angle=-90]{f5top.eps} \\
\includegraphics*[width=7.7cm,height=10.0cm,angle=-90]{f5bottom.eps} 

\caption{Representative color-color diagrams for {\em Spitzer} IRAC (top) and WISE (bottom)
         with colors of candidate YSOs shown. Source numbers correspond to Tables 3, 4, and 5. 
         The IRAC color-cut regions are
         from Gutermuth et al. (2008) and WISE regions are from Koenig et al. (2012).
         In the IRAC diagram, sources 16 and 26 lie in a region that includes both
         class I and heavily-reddened class II sources. The WISE colors of HH 199-IRS
         are shown for comparison but it was not detected as an X-ray source.
}
\end{figure}

\clearpage

\begin{figure}
\figurenum{6}
\epsscale{1.0}
\includegraphics*[width=12.0cm,height=15.53cm,angle=0]{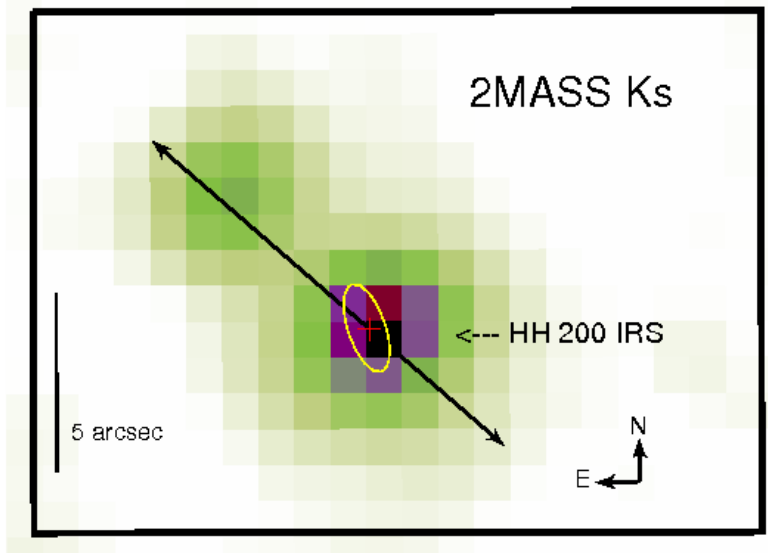}
\caption{2MASS K$_s$ (2.159 $\mu$m) image of HH 200-IRS =
         2MASS J205706.55$+$773655.94. The
         cross marks the radio counterpart position of source
         VLA 4 = J205706.703$+$773656.07 (Reipurth et al. 2004).
         The X-ray position error ellipse for CXO source 26 (J205706.72$+$773656.1)
         is overlaid and has
         semi-axes 1.$''$28 $\times$ 0.$''$52. The ellipse is  centered almost
         exactly on the VLA position. The arrows show the direction of
         the HH outflow axis at P.A. = 49$^{\circ}$/229$^{\circ}$ (Devine et al. 2009). 
         The second fainter source located 5.7$''$ NE of HH 200-IRS is
         2MASS J205707.89$+$773659.7 (J = 15.76, H = 14.24, K$_{s}$ = 13.44)
         and was not detected by {\em Chandra}.
}
\end{figure}

\clearpage
\begin{figure}
\figurenum{7}
\epsscale{1.0}
\includegraphics*[width=15.0cm,height=19.41cm,angle=0]{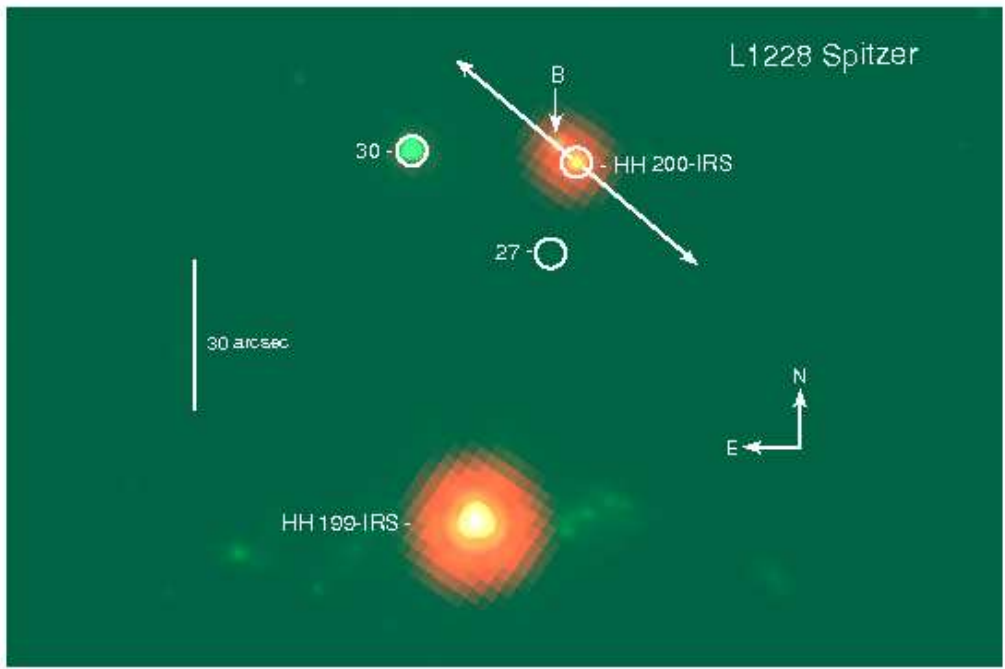}
\caption{Gaussian-smoothed Spitzer RGB image of the region near HH 199-IRS and HH 200-IRS.
 R = MIPS-24 (24 $\mu$m), G = I2 (4.5 $\mu$m), B = I1 (3.6 $\mu$m). The solid lines show the directions 
 of the HH 200 bipolar outflow. Circles mark the positions of Chandra X-ray sources (HH 200-IRS = CXO nr. 26). 
 The IR source marked B (= 2MASS J205707.89$+$773659.7) is located on the outflow 
 axis 5.7$''$ NE of HH 200-IRS but
 was not detected by {\em Chandra}. HH 199-IRS was not detected by Chandra. 
 Faint nebulosity associated with the HH 199 outflow
 is visible to the east and west of HH 199-IRS. }
\end{figure}

\clearpage
\begin{figure}
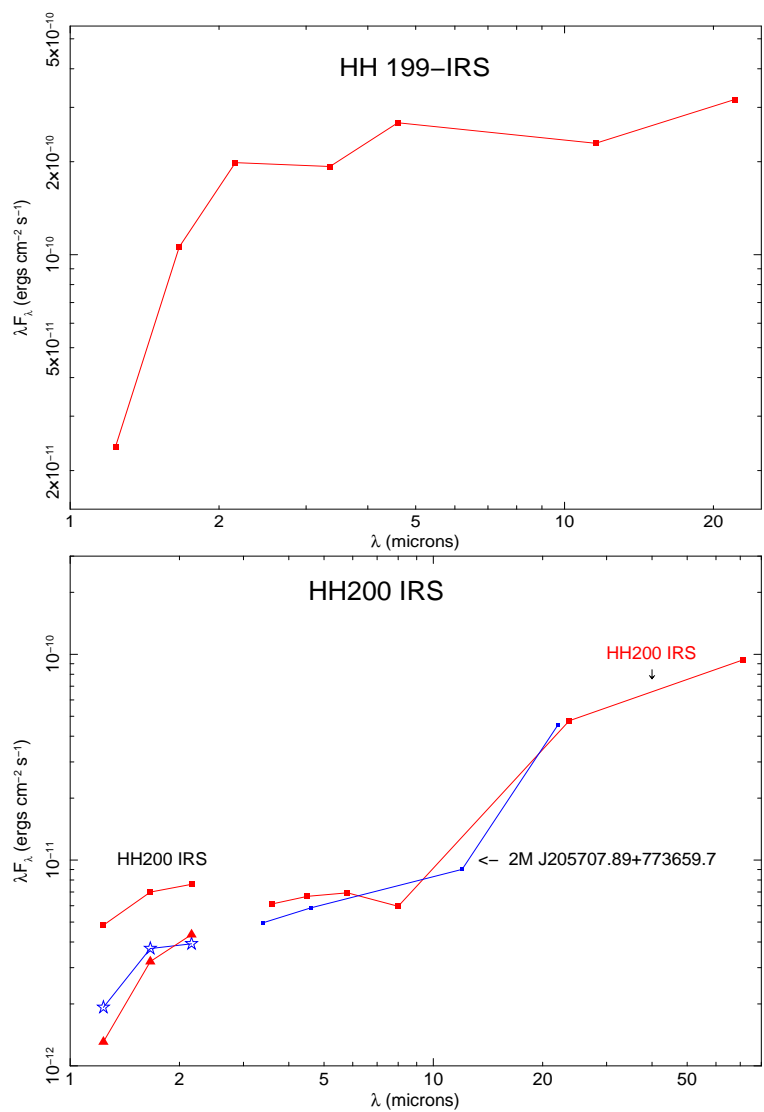

\figurenum{8}
\epsscale{1.0}
\includegraphics*[width=7.4cm,height=10.0cm,angle=-90]{f8top.eps} \\
\includegraphics*[width=7.4cm,height=10.0cm,angle=-90]{f8bottom.eps}
\caption{{\em Top}: IR SED of HH 199-IRS based on 2MASS and WISE photometry (Table 5).
         ~{\em Bottom}: IR SEDs of HH 200-IRS (red) and the close IR companion 2MASS J205707.89$+$773659.7
         (blue) located 5.$''$7 northeast of HH 200-IRS. The JHK$_{s}$ data for HH 200-IRS
         are from 2MASS (squares) and Chapman \& Mundy 2009 (triangles). The MIPS-70 $\mu$m
         data point is from c2d. The two sources are only partially resolved by MIPS at 24 $\mu$m
         and are unresolved by MIPS at 70 $\mu$m. The pair is not fully resolved by WISE.}
\end{figure}

\clearpage

\end{document}